\documentclass[usenatbib]{mn2e}
\usepackage{epsfig}
\usepackage{rotating}
\usepackage{multirow}

\bibpunct[, ]{(}{)}{;}{a}{}{,}

\title[WATs in ATLAS]{Wide-angle tail galaxies in ATLAS}
\author[Mao et al.]{Minnie~Y. Mao$^{1,2,3}$\thanks{e-mail: mymao@utas.edu.au},  Rob Sharp$^{2}$, D.~J. Saikia$^{3,4,5}$, Ray~P. Norris$^{3}$, 
\newauthor Melanie Johnston-Hollitt$^{6}$, Enno Middelberg$^{7}$ and Jim~E.~J. Lovell$^{1}$\\
$^{1}$School of Mathematics and Physics, University of Tasmania, Private Bag 37, Hobart, 7001, Australia\\
$^{2}$Anglo-Australian Observatory, PO Box 296, Epping, NSW, 1710, Australia\\
$^{3}$CSIRO Australia Telescope National Facility, PO Box 76, Epping, NSW, 1710, Australia\\
$^{4}$National Centre for Radio Astrophysics, Tata Insitute of Fundamental Research, Pune 411 007, India \\
$^{5}$ICRAR, University of Western Australia, Crawley, WA 6009, Australia \\
$^{6}$School of Chemical and Physical Sciences, Victoria University of Wellington, PO Box 600, Wellington, New Zealand\\
$^{7}$Astronomisches Institut, Ruhr-Universit\"at Bochum, Universit\"atsstr. 150, 44801 Bochum, Germany\\
}
\begin{document}

\date{200X}

\pagerange{\pageref{firstpage}--\pageref{lastpage}} \pubyear{200X}

\maketitle

\label{firstpage}

\begin{abstract}

  We present radio images of a sample of six Wide-Angle Tail (WAT)
  radio sources identified in the ATLAS 1.4\,GHz radio survey, and new
  spectroscopic redshifts for four of these sources. These WATs are in
  the redshift range of 0.1469$-$0.3762, and we find evidence of
  galaxy overdensities in the vicinity of four of the WATs from either
  spectroscopic or photometric redshifts. We also present follow-up
  spectroscopic observations of the area surrounding the largest WAT,
  S1189, which is at a redshift of $\sim$0.22.  The spectroscopic
  observations, taken using the AAOmega spectrograph on the AAT, show
  an overdensity of galaxies at this redshift.  The galaxies are
  spread over an unusually large area of $\sim$12 Mpc with a velocity
  spread of $\sim$4500 km s$^{-1}$. This large-scale structure
  includes a highly asymmetric FRI radio galaxy and also appears to
  host a radio relic. It may represent an unrelaxed system with
  different sub-structures interacting or merging with one another.
  We discuss the implications of these observations for future
  large-scale radio surveys. 
\end{abstract}


\section{Introduction}
\label{intro}

Wide-Angle Tail (WAT) galaxies are radio galaxies whose radio jets
appear to bend in a common direction. They are generally detected in
dynamical, non-relaxed clusters of galaxies \citep[e.g.][]{Burns90}
and may be used as probes or tracers for clusters \citep{Blanton00,
  Blanton01}. Clusters of galaxies are the largest gravitationally
bound structures in the Universe and are powerful testbeds of
cosmological models \citep[e.g.][] {Borgani04, Sahlen09,
  Kravtsov09}. Clusters also host diffuse radio emission in the form
of radio haloes and relics \citep{Giovannini00, Feretti05, Ferrari08,
  Giovannini09}.

The bent nature of WATs has commonly been attributed to strong
intra-cluster winds caused by dynamical interactions such as
cluster-cluster mergers \citep{Burns98}. WATs are preferentially found
in enhanced X-ray regions \citep{Pinkney00} and are usually associated
with dominant cluster galaxies \citep{Owen76}. \citet{Mao09a} found
the tailed radio galaxies, including WATs, to be located in the
densest regions of clusters in the local Universe, consistent with
earlier studies \citep[e.g.][] {Burns90, Blanton00, Blanton01}. Thus WATs 
represent valuable tracers of high density regions in the intracluster 
medium (ICM), and this approach has been used in a number of recent studies
\citep[e.g.][] {Blanton00, Blanton03, Smolcic07, Giacintucci09, 
Kantharia09, Oklopcic10}.

Here we present the radio properties of six WATs that we have
identified in ATLAS, the Australia Telescope Large Area Survey,
carried out with the Australia Telescope Compact Array (ATCA) at
1.4\,GHz \citep{Norris06,Middelberg08}.  ATLAS
\footnote{http://www.atnf.csiro.au/research/deep/index.html} will
image seven square degrees of sky over two fields to an rms
sensitivity of 10 $\mu$Jy beam$^{-1}$. The ATLAS fields have been
observed with a number of different ATCA configurations, and the
typical resolution of the observations is $\sim$10 arcsec.  The two
ATLAS fields, Chandra Deep Field South (CDFS) and European Large Area
ISO Survey-South 1 (ELAIS-S1), were chosen to coincide with the
\textit{Spitzer} Wide-Area InfraRed Extragalactic (SWIRE) survey
program \citep{Lonsdale03} so that corresponding optical and infrared
photometric data are available.

In addition to the radio properties we present new spectroscopic redshifts
for four of the WATs and follow-up spectroscopic observations of galaxies
in the vicinity of the largest WAT in order to probe its surrounding
structure. This WAT was first identified as radio source S1189 by
\citet{Middelberg08}, and is associated with the SWIRE source
SWIRE4\_J003427.54-430222.5 \citep{Lonsdale03}. 

In this paper we present a summary of the data in Section 2, while the
WATs in ATLAS are presented in Section 3. Section 4 presents the
results of spectroscopic observations of S1189 and its surrounding
region, and discusses the large-scale structure in its vicinity.  In
Section 5 we discuss cosmological inverse-Compton quenching and the
implications for deep wide radio surveys and ATLAS. This paper uses
H$_0$ = 71 km s$^{-1}$ Mpc$^{-1}$, $\Omega$$_M$ = 0.27 and
$\Omega$$_\Lambda$ = 0.73 and the web-based calculator of
\citet{Wright06} to estimate the physical parameters. Vega magnitudes
are used throughout.
   
\section{Data}
\subsection{Radio Data}
ATLAS radio observations are currently partially complete with an rms
noise of $\sim$20 - 30 $\mu$Jy beam$^{-1}$ at 1.4 GHz. The data used
in this paper are taken from the first ATLAS catalogues
\citep{Norris06,Middelberg08} which contain 2004 radio sources. We
expect $\sim$16000 radio sources at the completion of the survey.

\subsection{Spectroscopy}\label{specdata}
As part of ATLAS, we are undertaking a program of redshift
determination and source classification of all ATLAS radio sources
with AAOmega \citep{Sharp06} on the Anglo-Australian Telescope
(AAT). We are currently partway through our ATLAS spectroscopy
campaign. A summary of these observations are presented by
\citet{Mao09b} while the detailed results will be presented by Mao et
al. (in preparation). 169 ATLAS sources already have spectroscopic
redshifts from the literature. We have obtained 395 new spectroscopic
redshifts using AAOmega giving a total so far of 564 spectroscopic
redshifts: 261 in CDFS and 303 in ELAIS-S1. All of the WATs 
presented in this paper have spectroscopic data from either our AAT
observations or 2dFGRS \citep{Colless01}.

\subsection{Follow-up Spectroscopy of the region around S1189}\label{service}
\begin{figure}
\begin{center}
\includegraphics[angle=-90, scale=0.35]{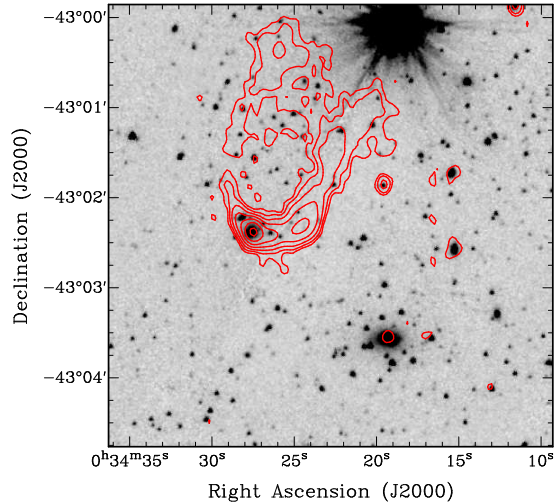}
\end{center}
\caption{SWIRE 3.6-$\mu$m image of the WAT, S1189, and the putative cD
  galaxy located south-west of the WAT. The 1.4 GHz radio contours
  which are overlaid start from 100 $\mu$Jy beam$^{-1}$ (3 $\times$
  rms) and increase by factors of 2. }\label{S1189cD}
\end{figure}

There appears to be a cluster of galaxies within $\sim$2 arcmin of
S1189 in the optical and infrared images (see Fig. \ref{S1189cD}). We
obtained AAOmega observations for sources within a degree of S1189 in
service mode during the night of 2008 October 18.  The AAOmega
spectrograph was used in multi-object mode \citep{Saunders04, Sharp06}
and centred on the WAT. We used the dual beam system with the 580V and
385R Volume Phase Holographic (VPH) gratings centred at $\lambda$4800
and $\lambda$7150 covering the spectral range between 3700\AA\ and
8500\AA\ at central resolutions in each arm of R$\sim$1300 per 3.4
pixel spectral resolution element. The 5700\AA\ dichroic beam splitter
was used. Observing conditions were good with clear skies and an
average seeing of $\sim$1.6 arcsec. Two fibre configurations were
observed with 3 $\times$ 1200 sec integrations and associated
quartz-halogen flat fields and combined CuAr+FeAr, Helium and Neon arc
lamp frames.
 
Targets were identified from the SWIRE catalogues
\citep{Lonsdale03}. The target magnitude range was limited to 19 $<$ R
$<$ 20.5. The bright limit was chosen to select against foreground
galaxies based on the expected low number of galaxies brighter than L*
in the potential cluster. The faint limit was chosen due to the
bright-of-moon service observations. The magnitude range yielded
$\sim$7000 sources within a one degree radius (the field of view of
the 2dF/AAOmega fibre positioner \citep{Lewis02}) centered on
S1189\footnote{The SWIRE input catalogue of \citet{Lonsdale03}
  excludes a number of small regions at the outer edge of the
  field.}. Targets were prioritized based on radial separation from
the WAT S1189 with the exception of the putative cD galaxy which was
assigned the highest priority to ensure that its redshift was
obtained. Targets farther than 5 arcmin were randomly sampled using
Fisher-Yates shuffles, to decrease the input catalogue to a practical
working sample for the \textsc{configure} software and the Simulated
Annealing fibre allocation algorithm \citep{Miszalski06}, as given in
Table \ref{priority}. Regrettably no star-galaxy separation was
performed resulting in the inclusion of stars in the input catalogue.
Although $\sim$400 AAOmega science fibres are available, fibre
allocation requires target separations in excess of 30 arcsec due to
physical limitations. Consequently two independent fibre
configurations were observed to secure as many high priority sources
as possible.

Data reduction followed the standard pattern for AAOmega spectroscopy
using the \texttt{2dfdr} software package. The red and blue arms were
reduced independently and then spliced together so as to produce a
continuous spectrum. The redshift was then determined from the spectra
using \texttt{runz}.

\begin{table*}
\begin{center}
\caption{Priority assignment for the target observations. AAOmega is configured based on source location and user-defined priority assignment with 9 being the highest priority and 1 being the lowest. Columns 1 and 2 list the priority assignment and the number of sources in each priority bin. Column 3 presents the number of sources for which we were able to obtain redshifts, while Column 4 gives the radii of the priority bin from S1189. Column 5 describes how many sources were selected randomly using Fisher-Yates shuffles. Priority 1 sources were not included in the target list. }\label{priority}
\begin{tabular}{lllll}
\\
\hline
Priority & No. sources & Redshifts & Radii & Comment\\
\hline
9 & 1 & 1 & & putative cD\\
8 & 12 & 0 & $<$ 2$^\prime$& \\
7 & 60 & 7 & 2$^\prime$ to 5$^\prime$& \\
6 & 100 & 8 & 5$^\prime$ to 10$^\prime$ & 100/187 randomly selected\\
5 & 200 & 15 & 10$^\prime$ to 15$^\prime$ & 200/328 randomly selected\\
4 & 200 & 28 & 1$^\prime$5 to 30$^\prime$ & 200/1956 randomly selected\\
3 & 200 & 19 & 30$^\prime$ to 1 deg & 200/4426 randomly selected\\
2 & 46 & 8 & & sources with previously determined z$_{spec}$\\
1 & 6197 & 0 & $>$ 5$^\prime$ & sources not randomly selected\\
\hline
\end{tabular}
\end{center}
\end{table*}

\section{WATs in ATLAS}
\begin{figure*}
\begin{center}
\begin{tabular}{cccc}

 \multirow {10}{*}{S132} & \includegraphics[angle=-90,scale=0.17]{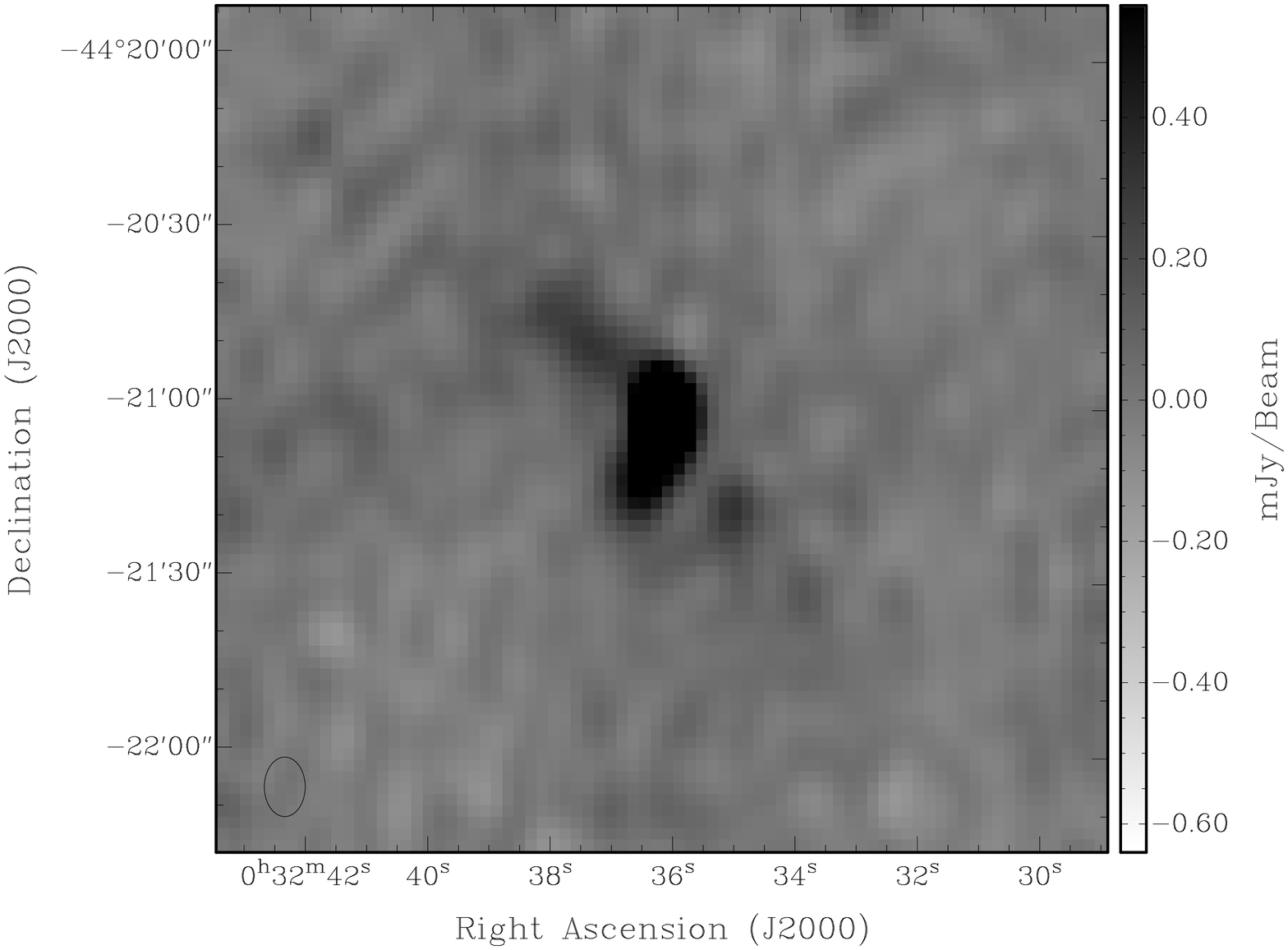} & \includegraphics[angle=-90,scale=0.17]{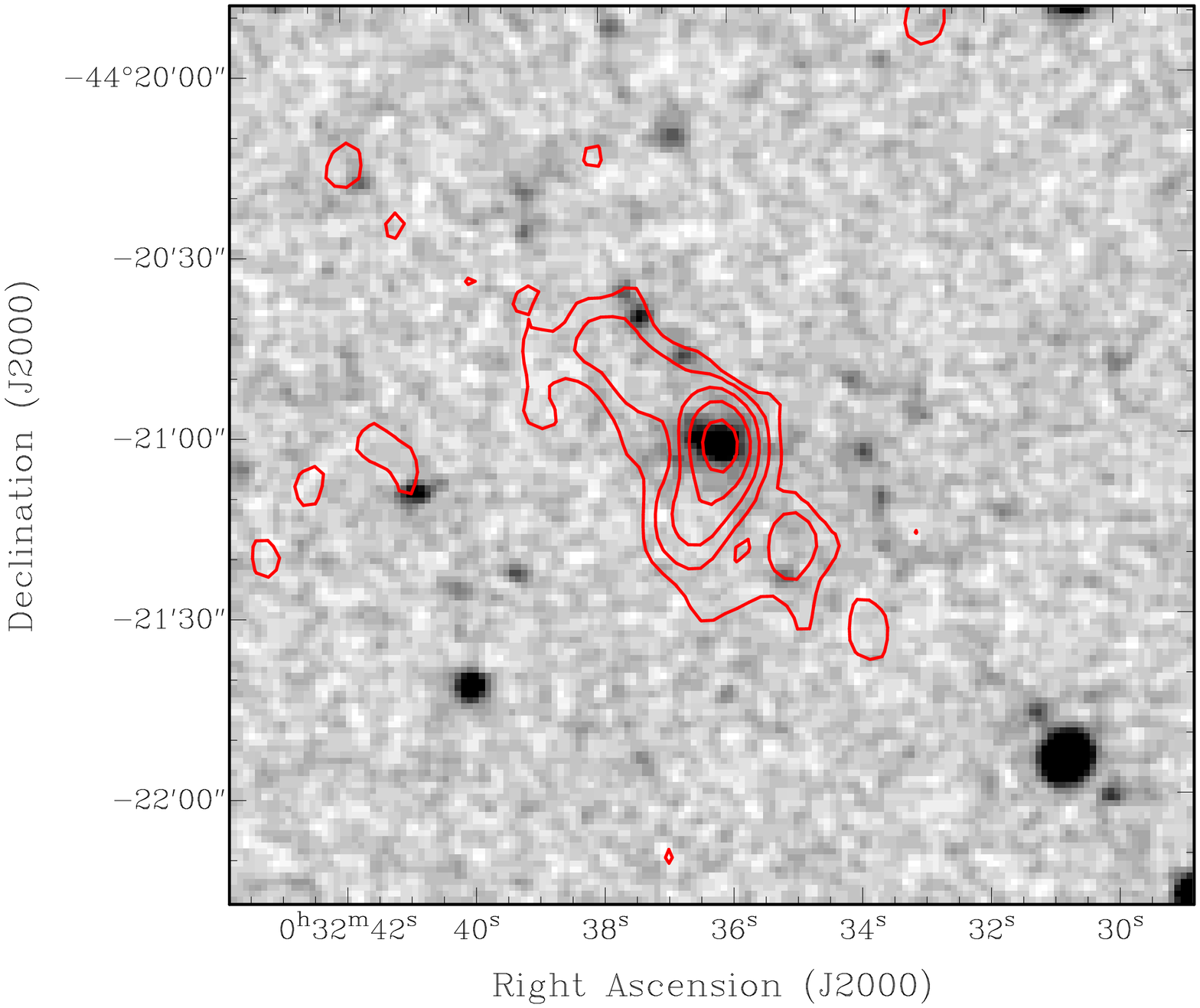} & \includegraphics[angle=-90,scale=0.17]{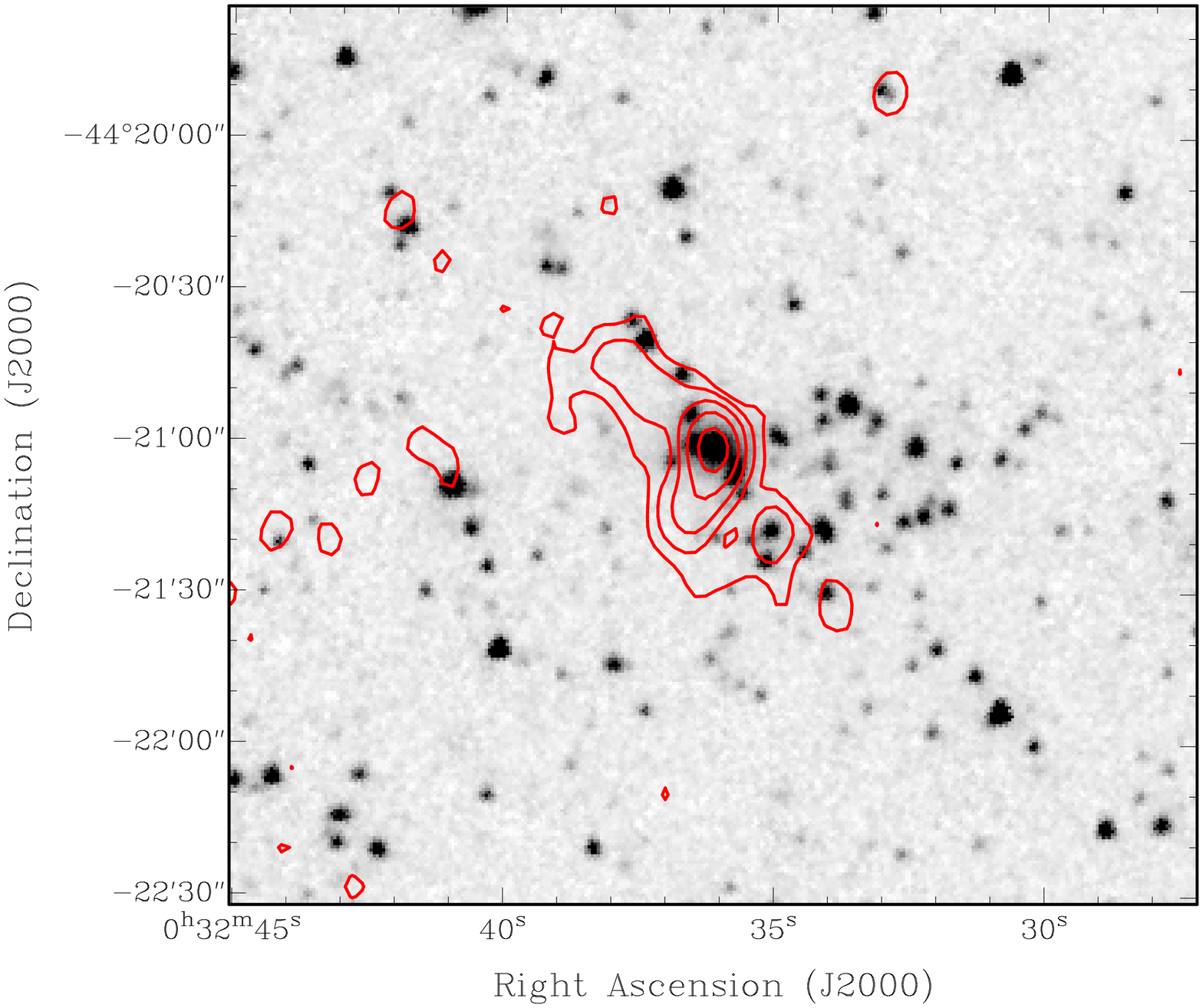}\\
  \multirow {10}{*}{S483} & \includegraphics[angle=-90,scale=0.17]{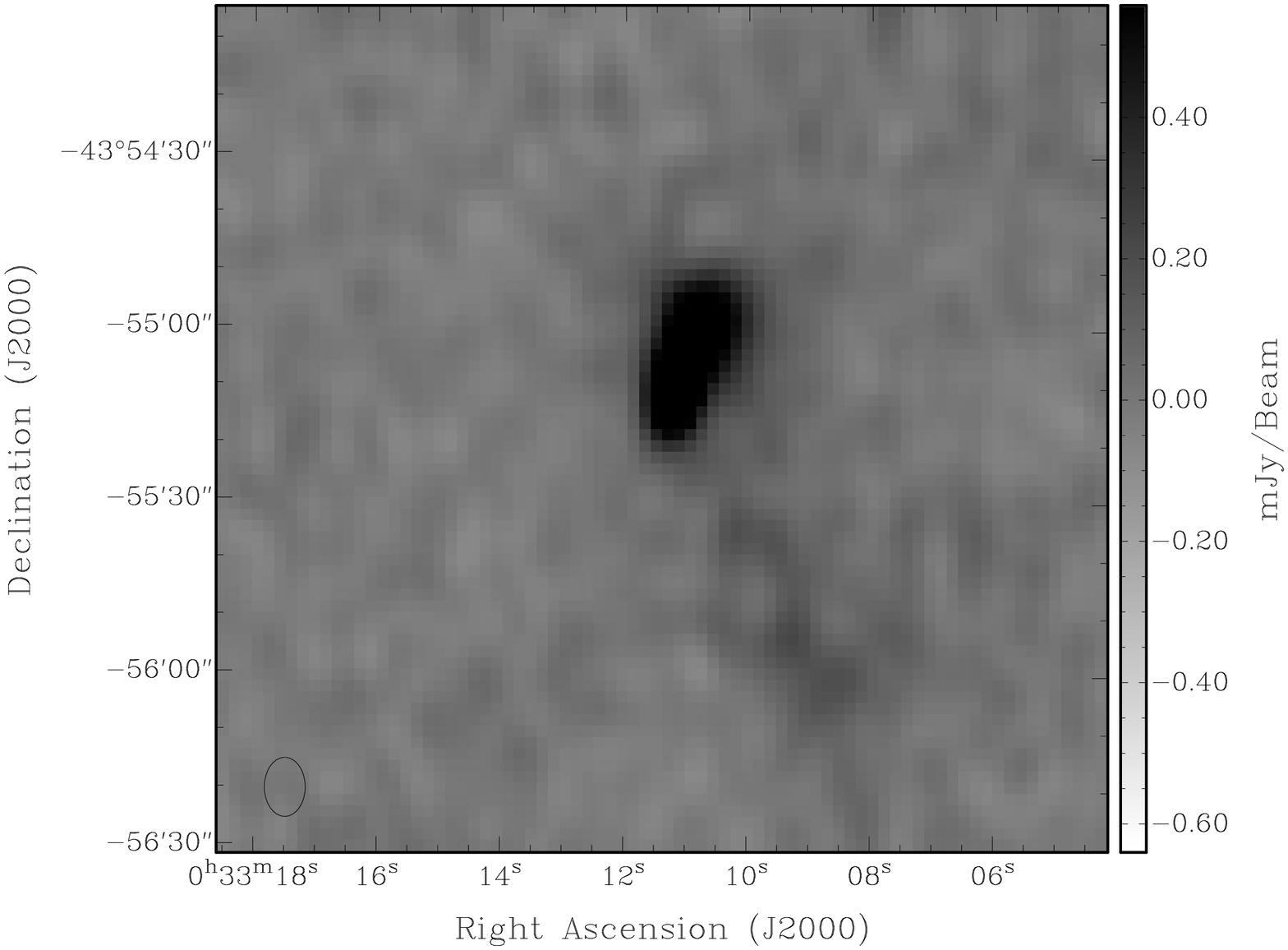}\ & \includegraphics[angle=-90,scale=0.17]{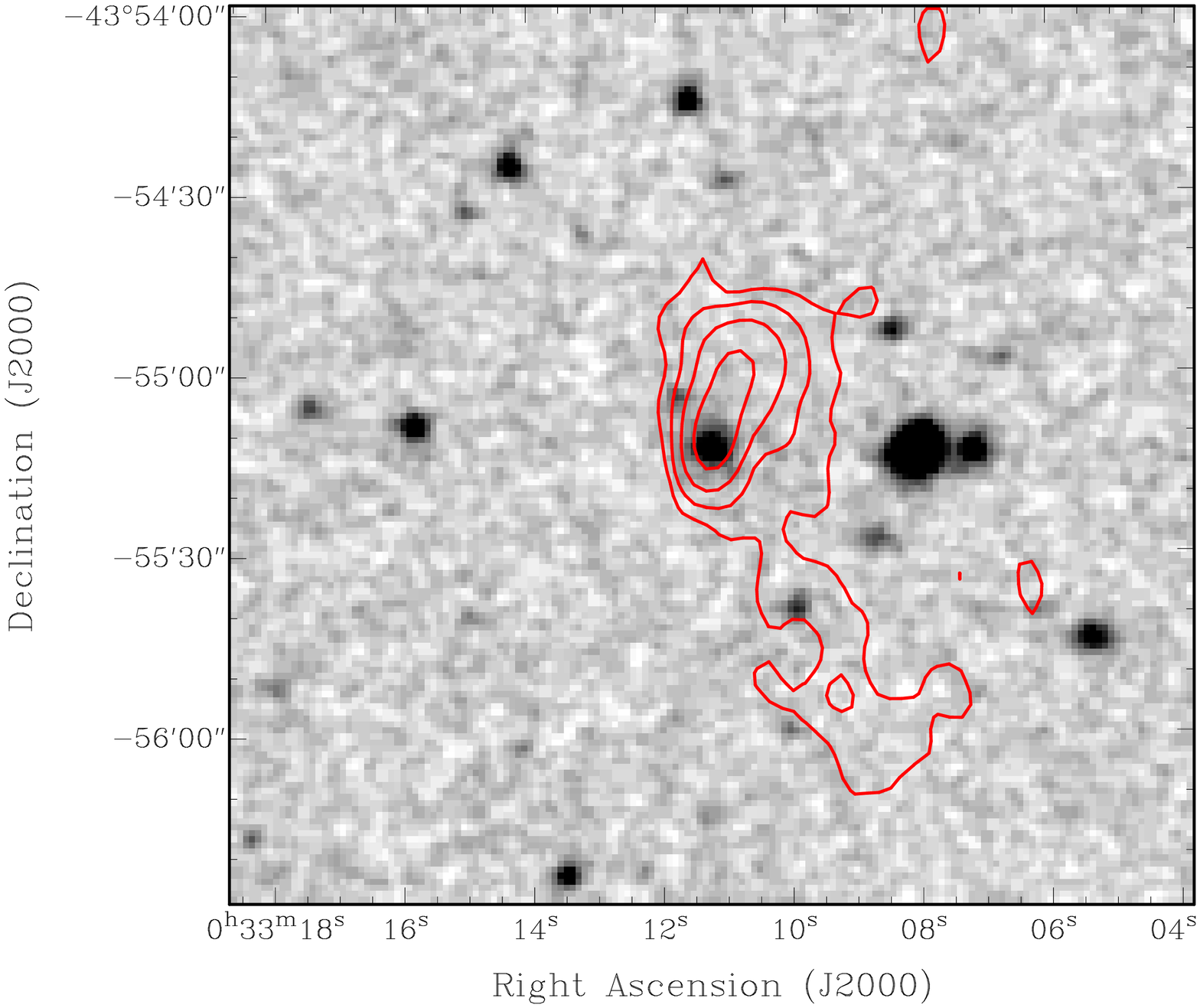} & \includegraphics[angle=-90,scale=0.17]{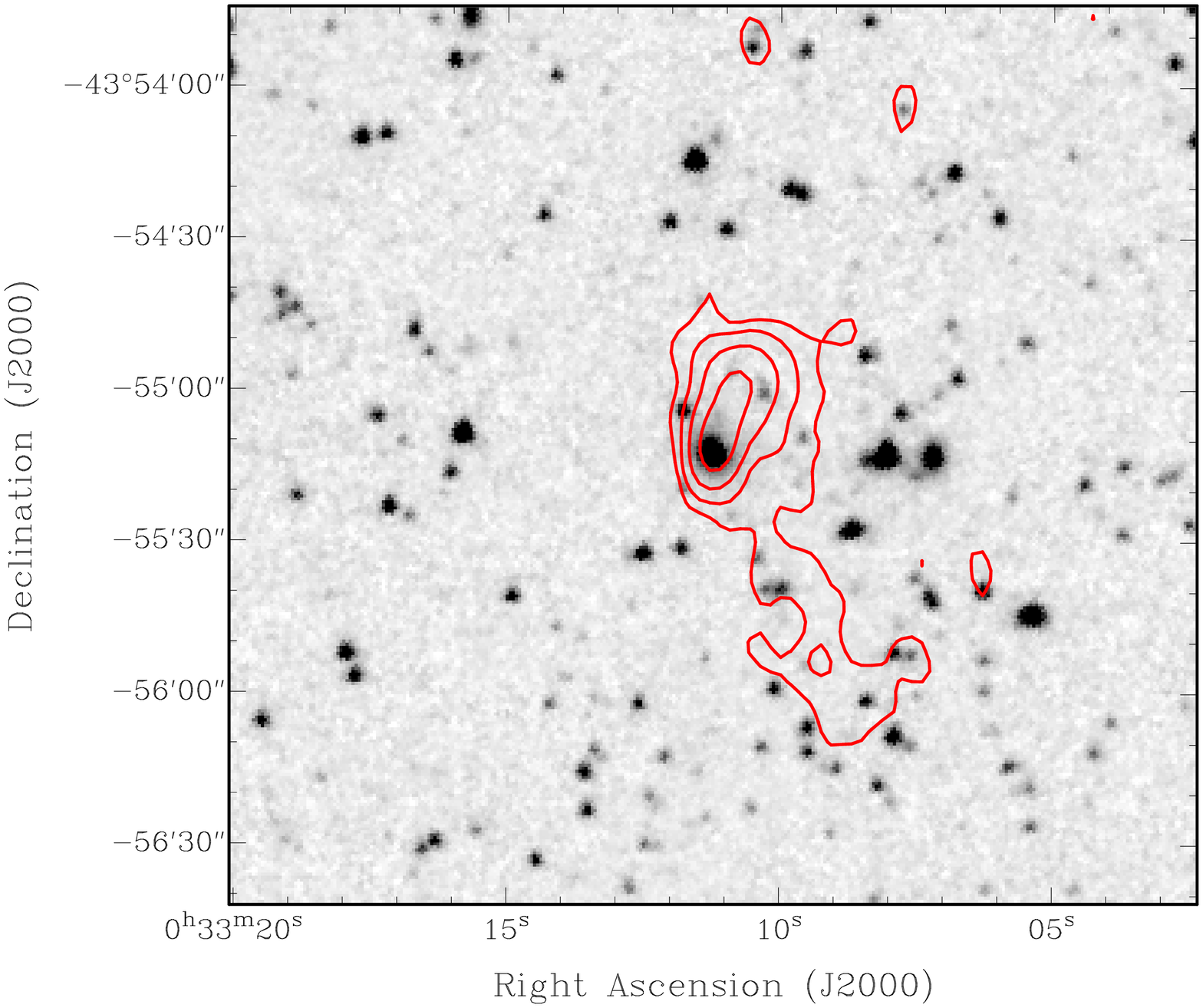}\\
  \multirow {10}{*}{S1189} & \includegraphics[angle=-90,scale=0.17]{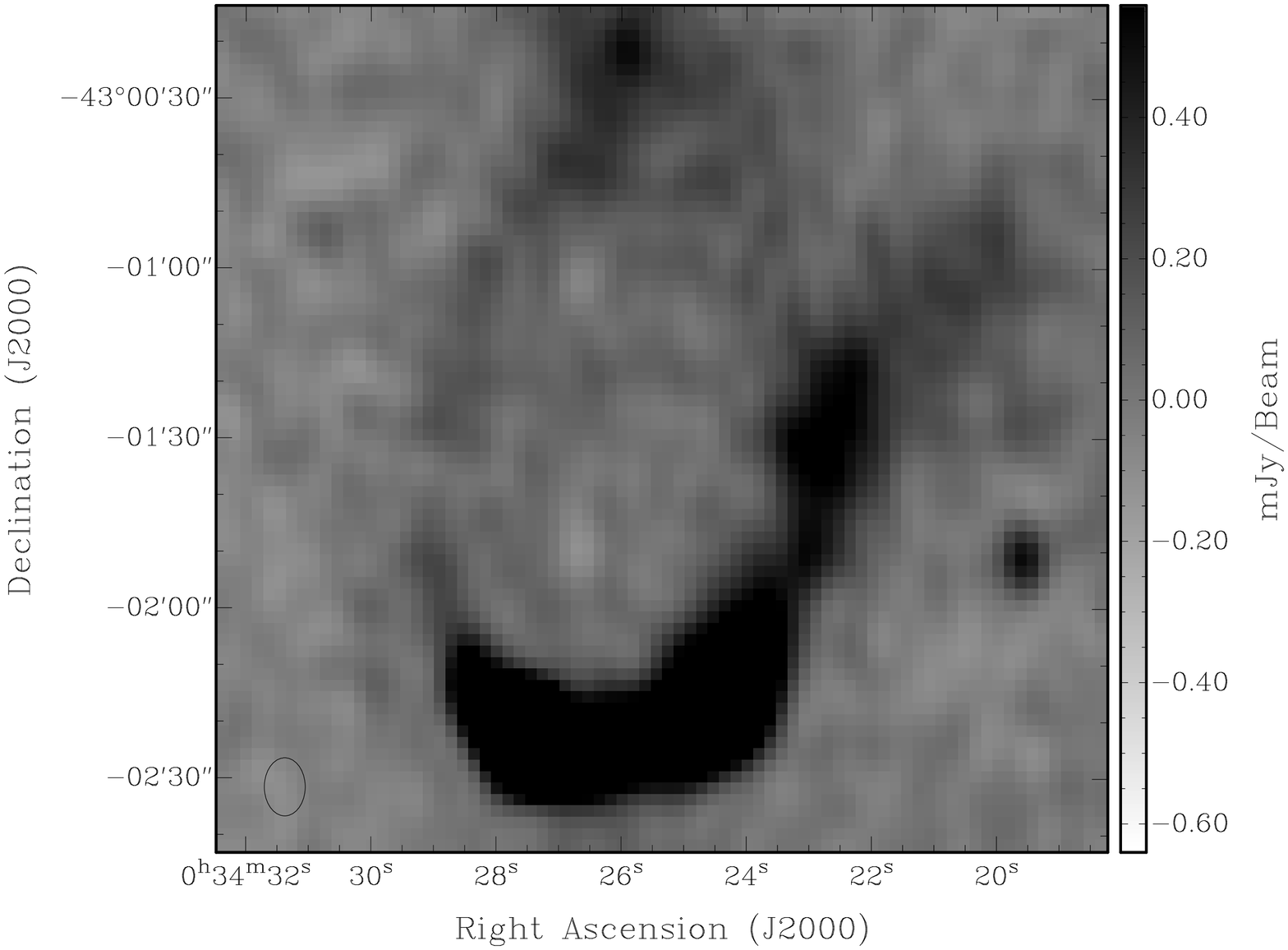}& \includegraphics[angle=-90,scale=0.17]{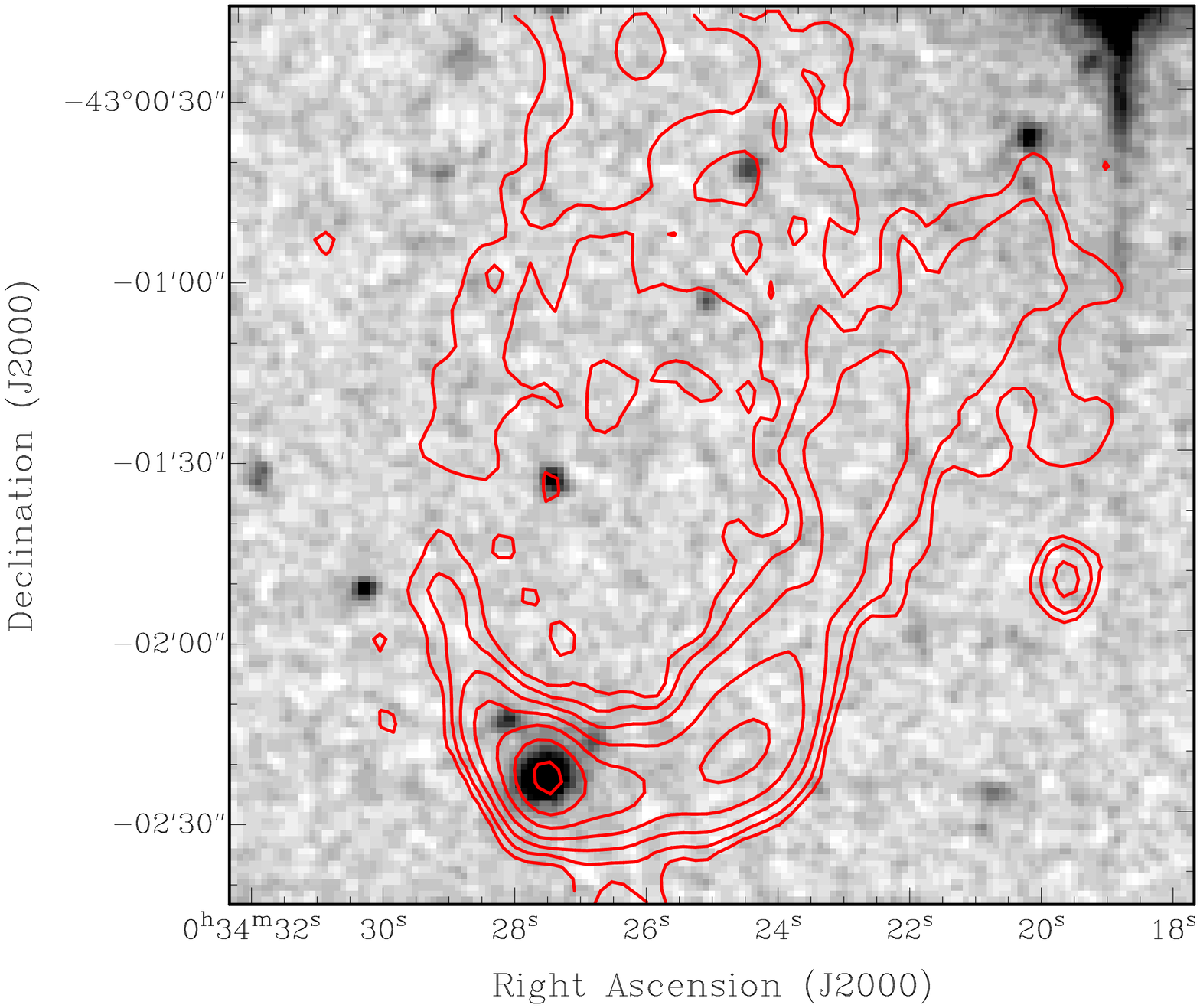} & \includegraphics[angle=-90,scale=0.17]{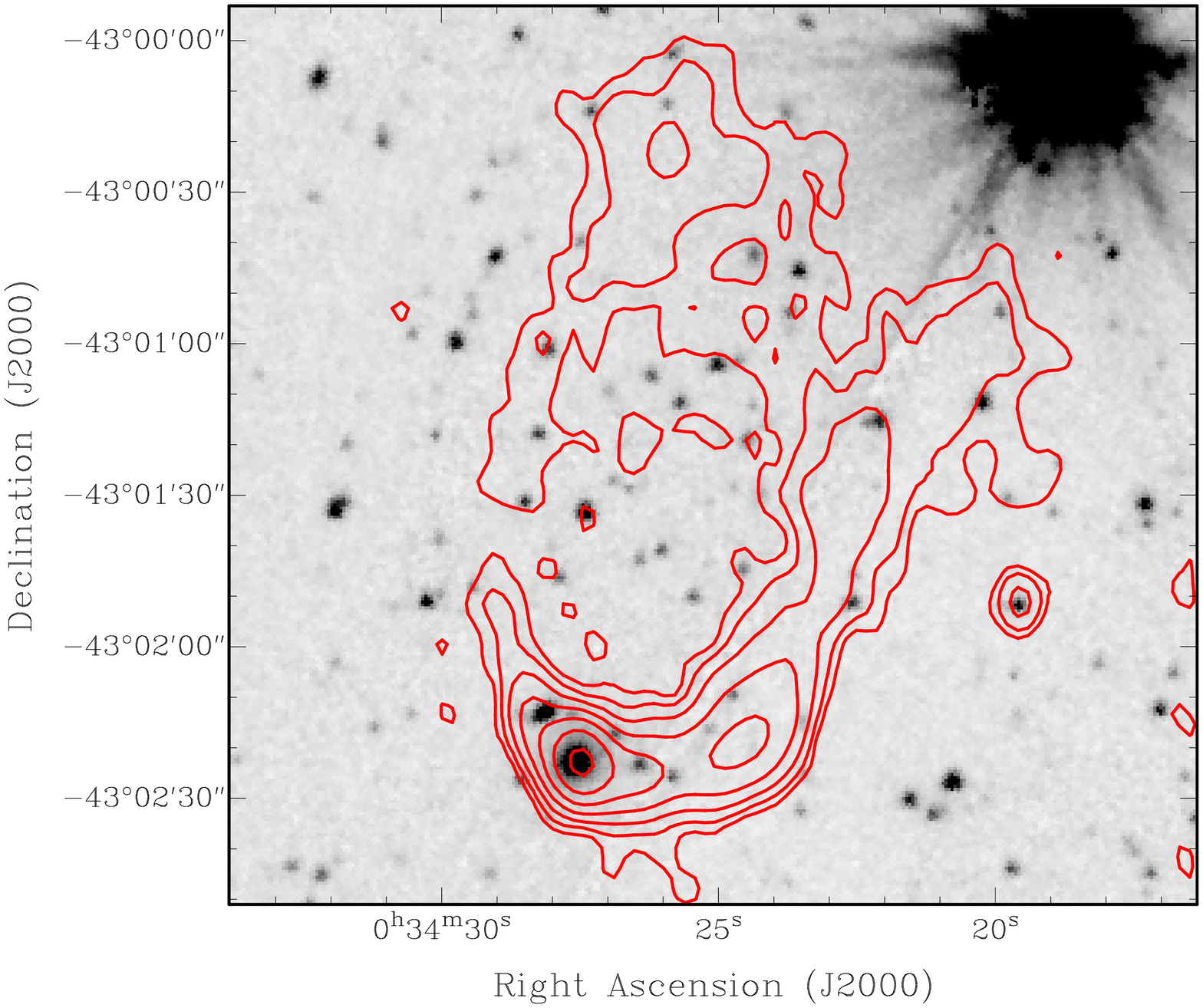}\\
 \multirow {10}{*}{S1192} & \includegraphics[angle=-90,scale=0.17]{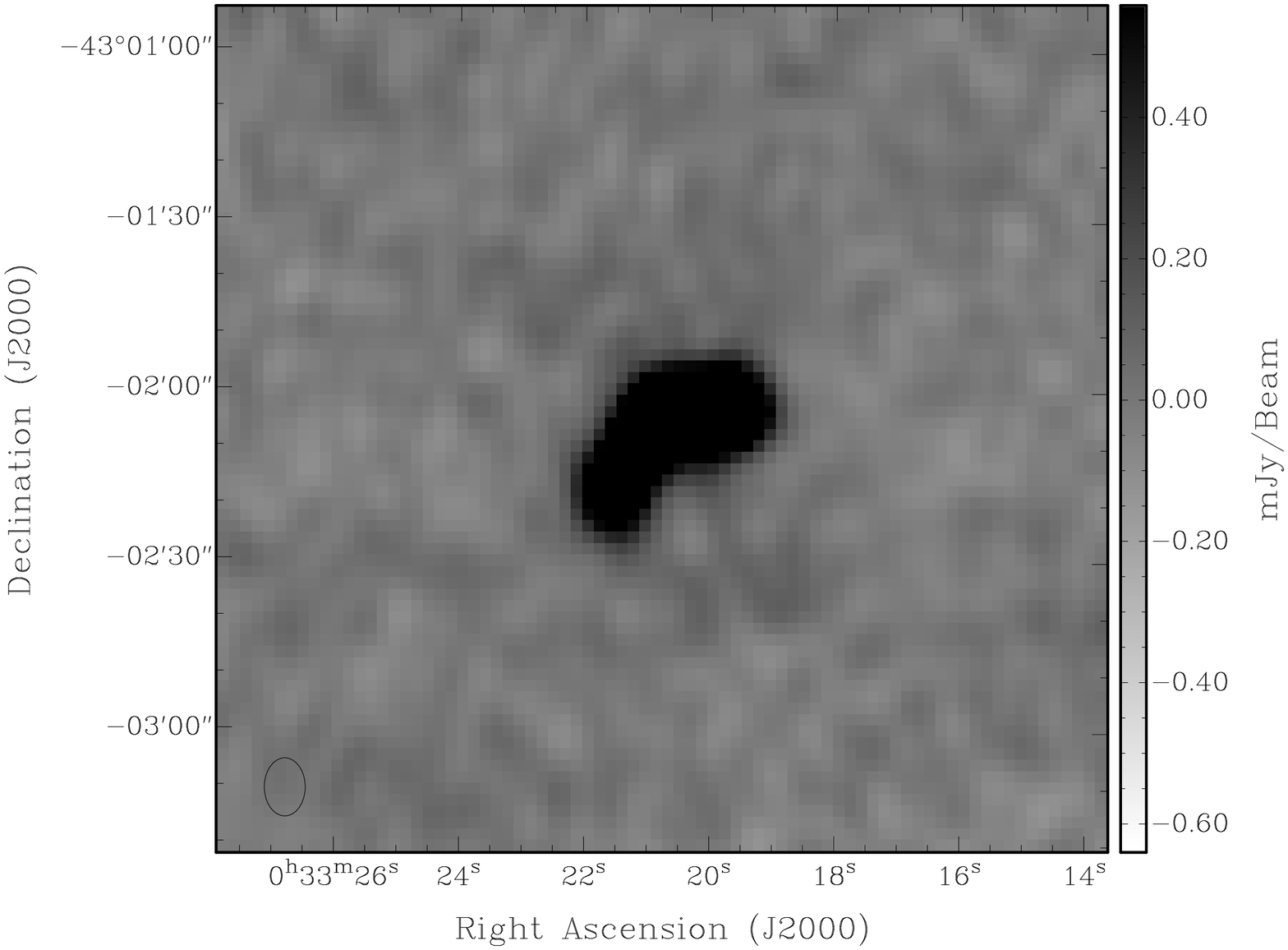}& \includegraphics[angle=-90,scale=0.17]{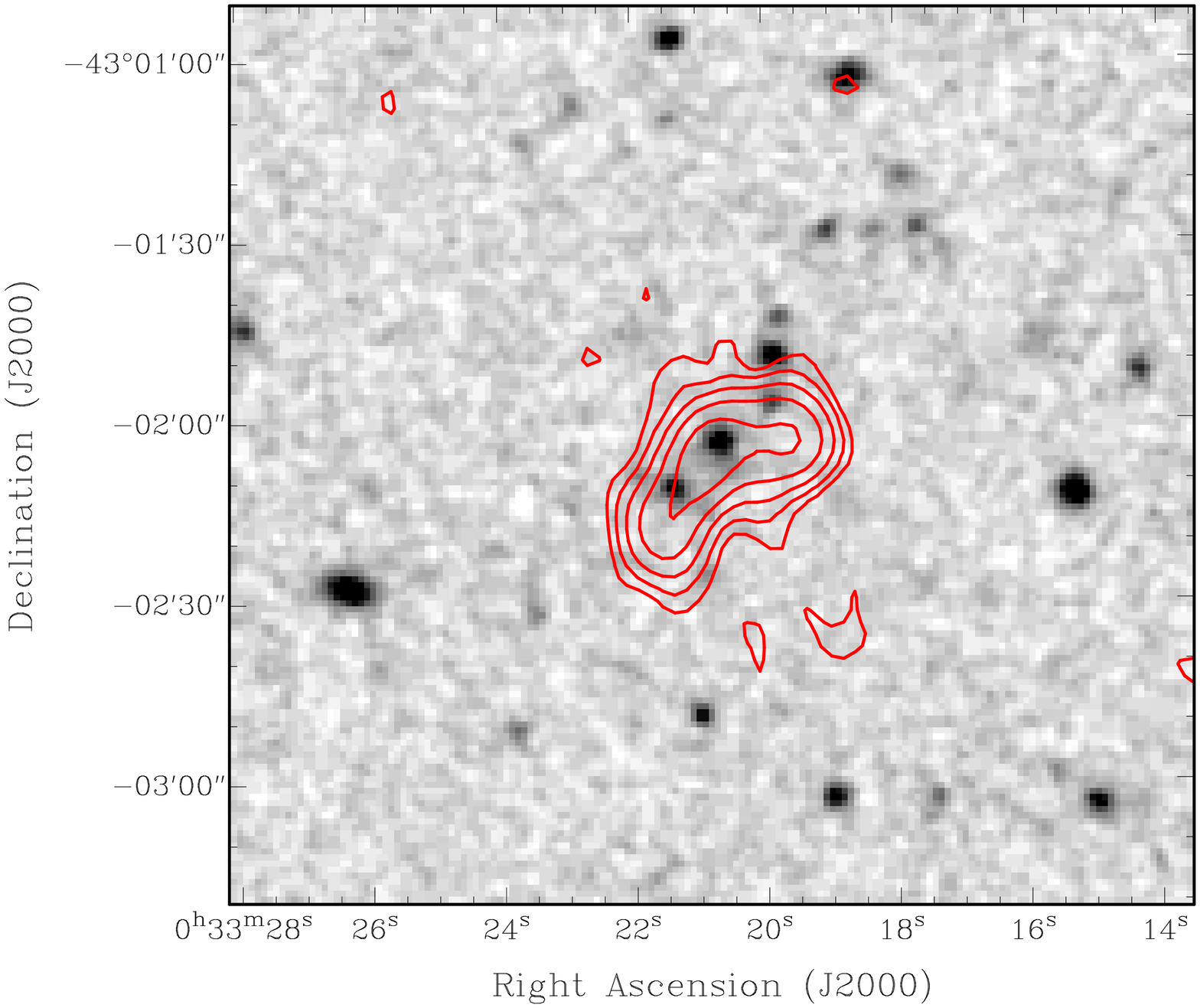} & \includegraphics[angle=-90,scale=0.17]{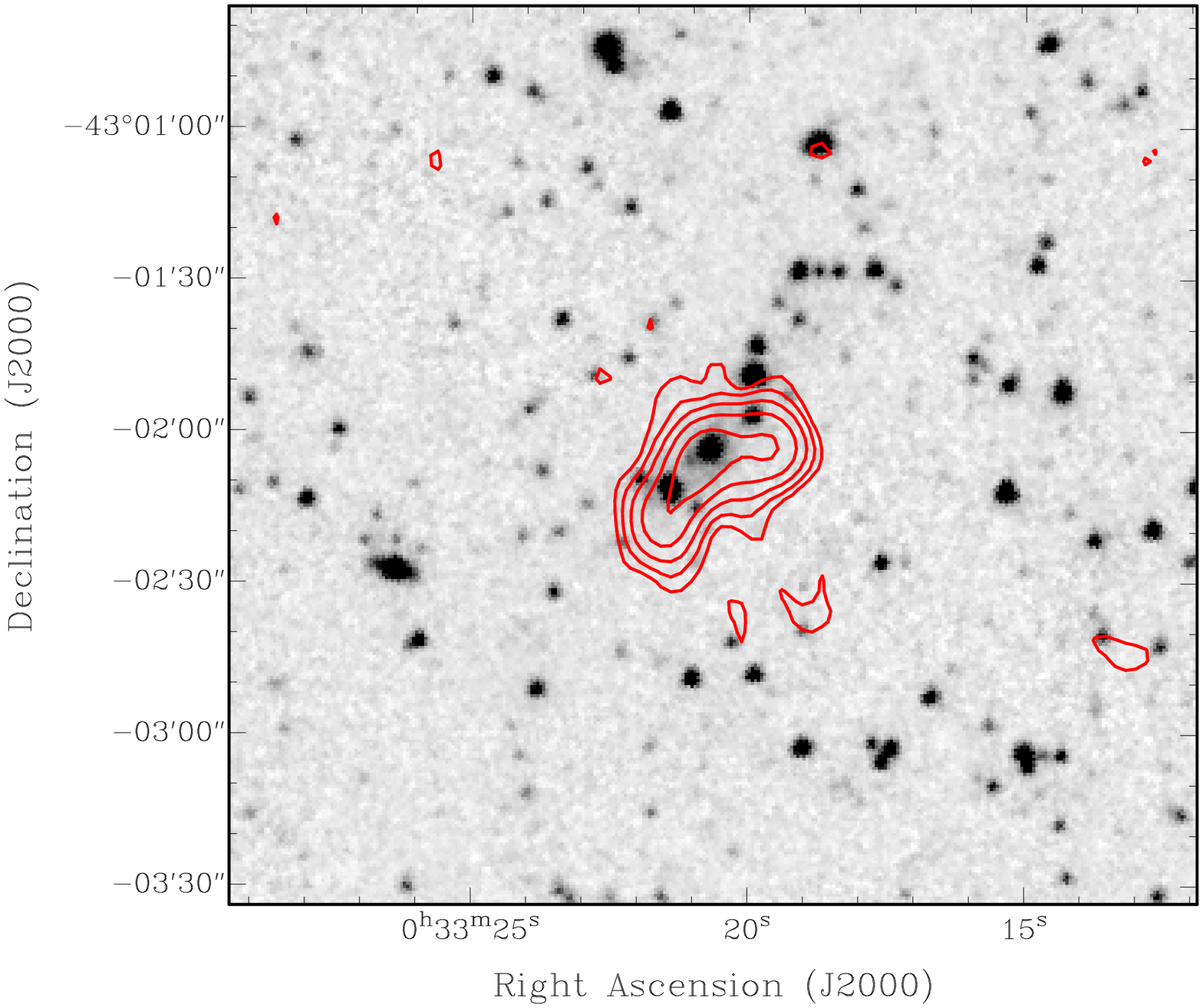} \\
 \multirow {10}{*}{S031} &  \includegraphics[angle=-90,scale=0.17]{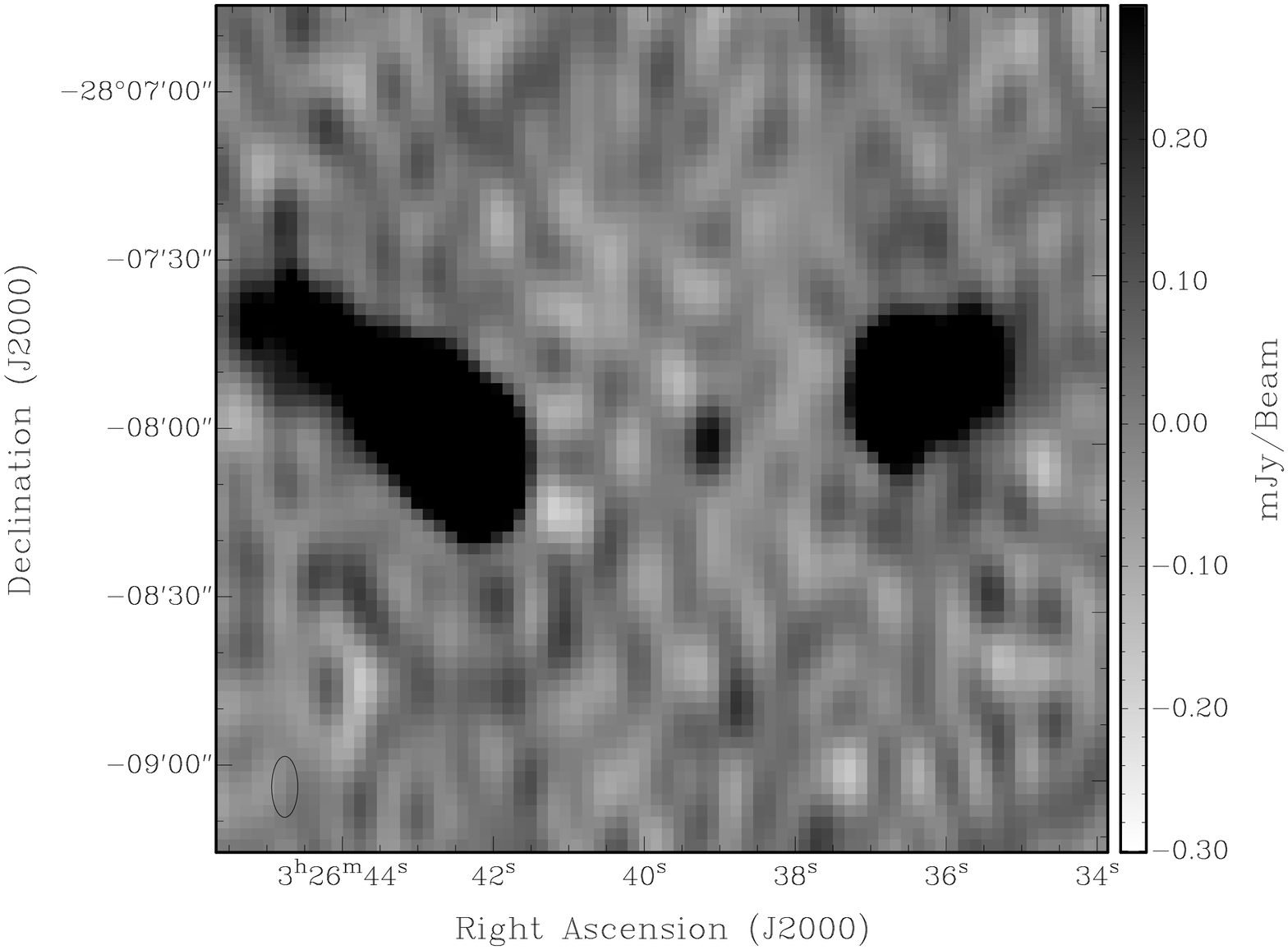}& \includegraphics[angle=-90,scale=0.17]{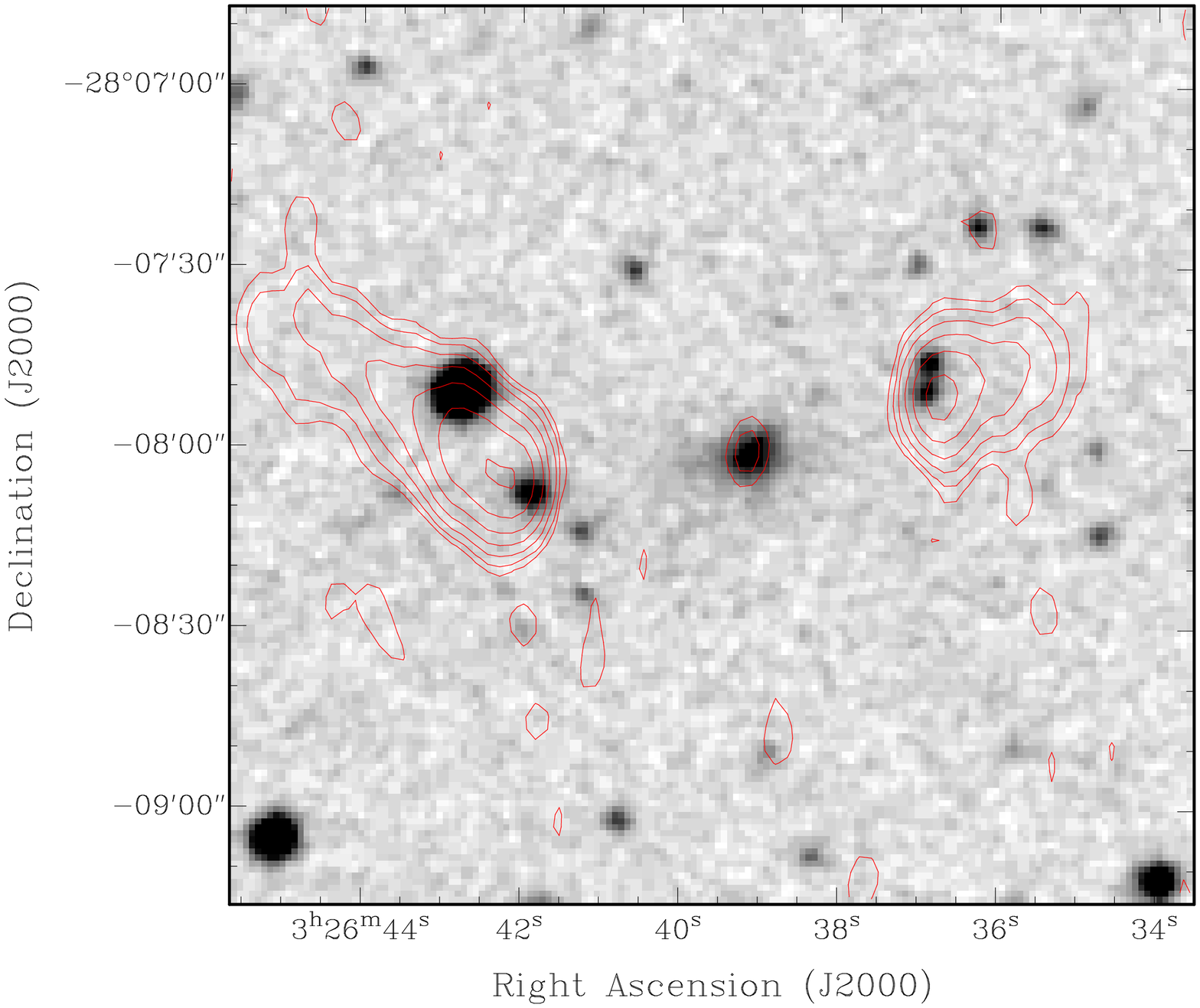} & \includegraphics[angle=-90,scale=0.17]{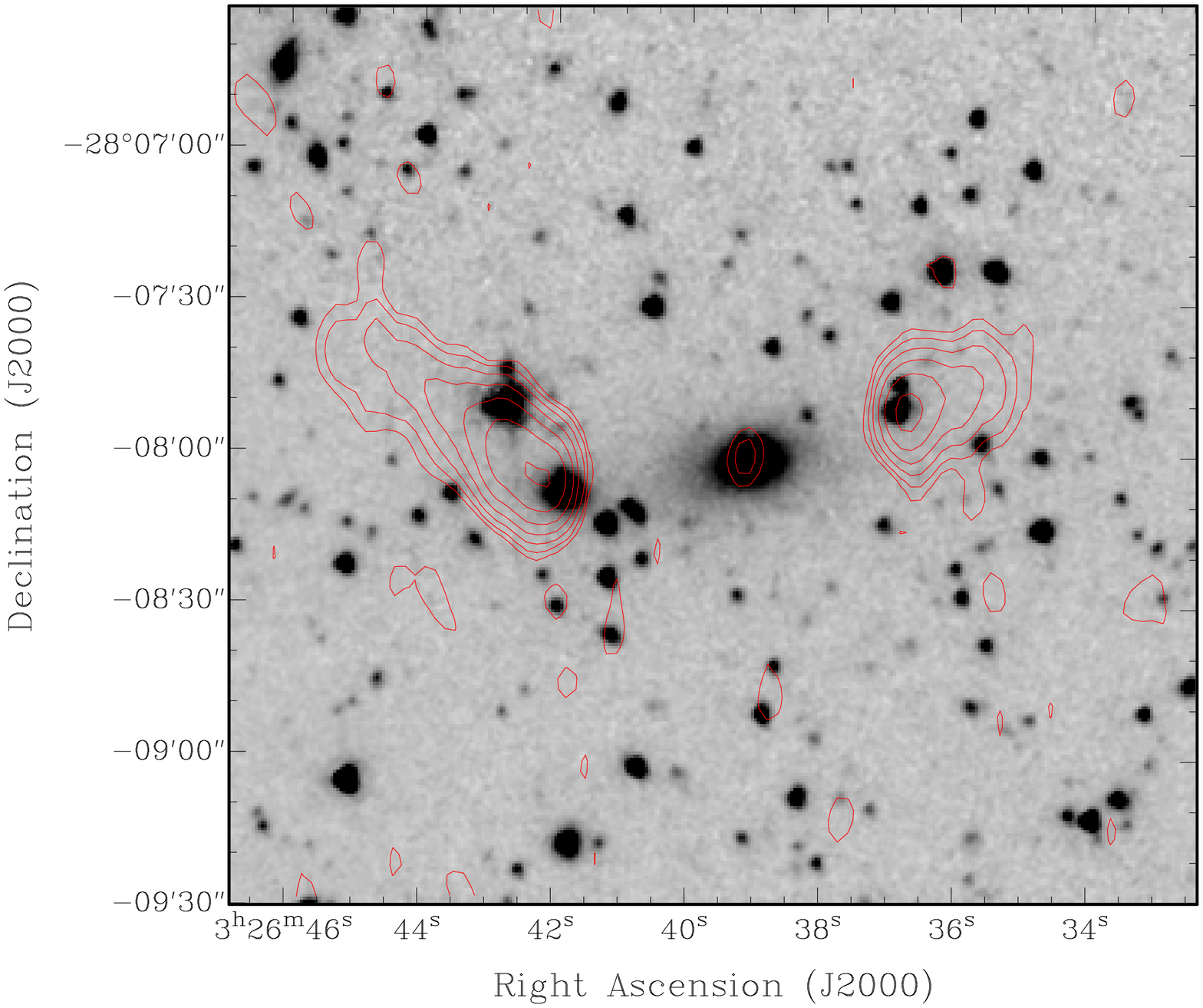}\\
 \multirow {10}{*}{S409} & \includegraphics[angle=-90,scale=0.17]{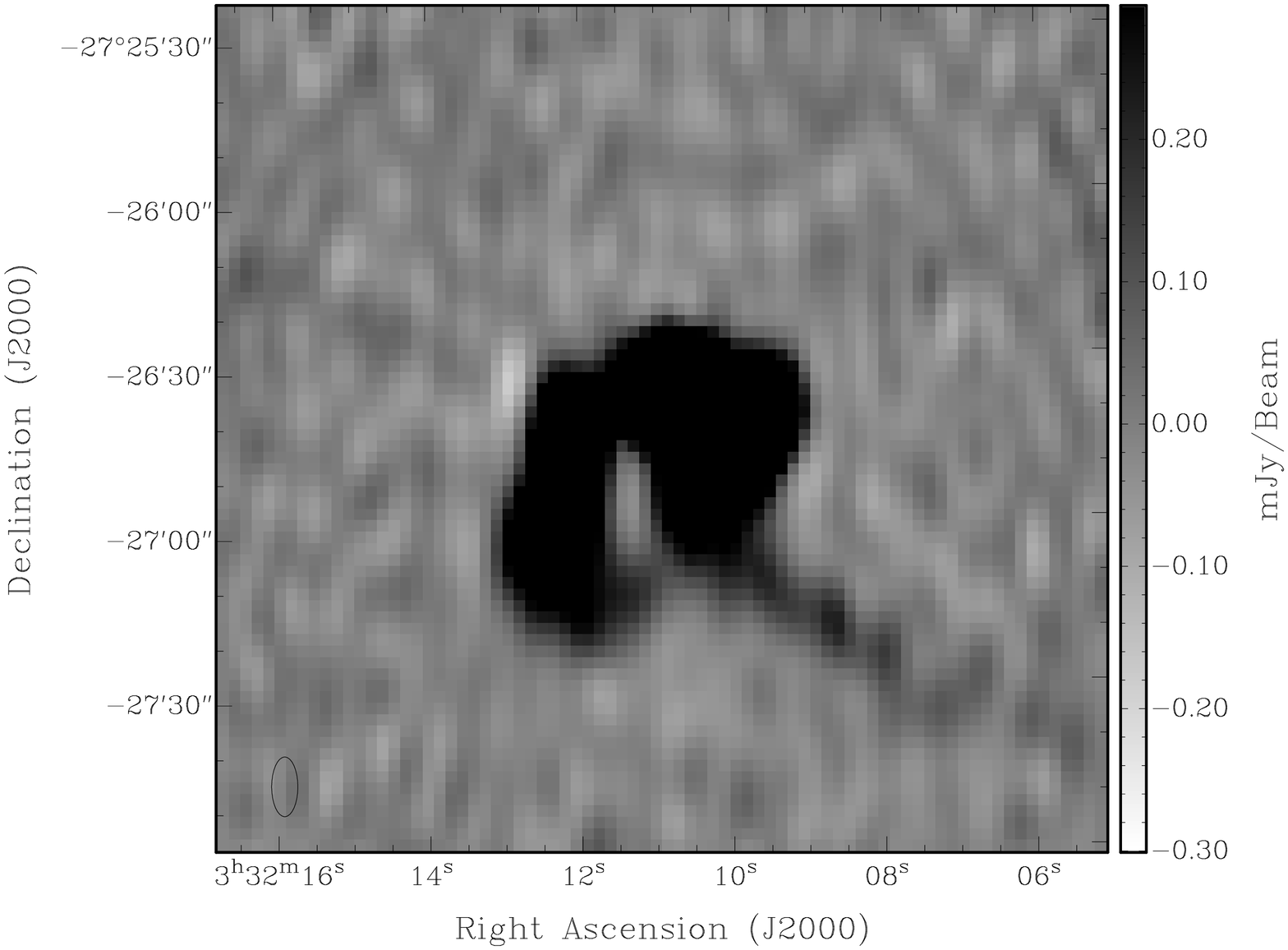}& \includegraphics[angle=-90,scale=0.17]{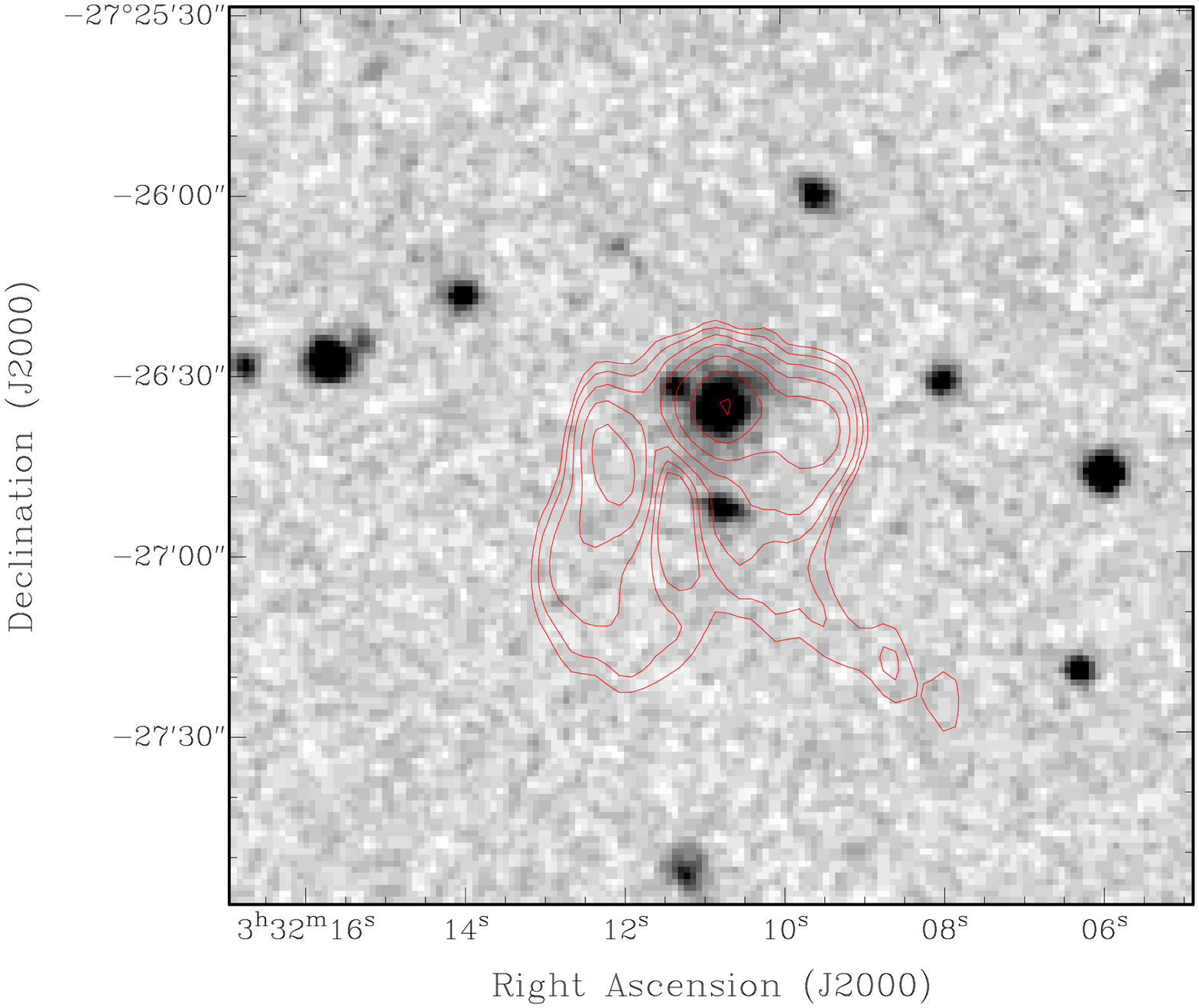} & \includegraphics[angle=-90,scale=0.17]{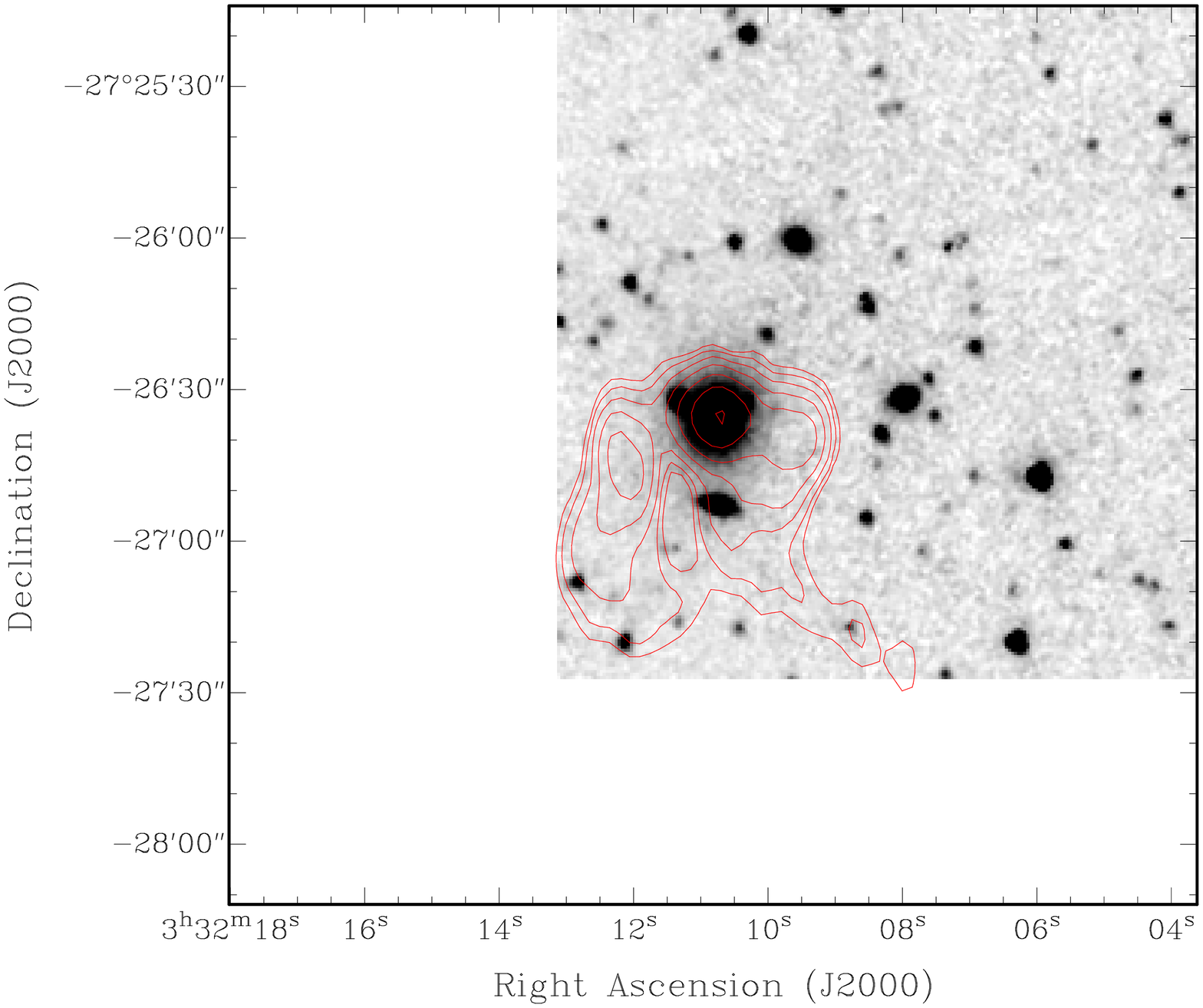}\\

\end{tabular}
\end{center}
\caption{The six WATs in ATLAS. From top to bottom the WATs are S132, S483, S1189 and S1192 in ELAIS-S1, and S031 and S409 in CDFS.  The left column shows the 1.4 GHz radio continuum emission of the WATs in greyscale.The middle column shows the radio contours overlaid on DSS red images. The right column shows the radio contours overlaid on 3.6-$\mu$m IRAC images. The radio contours start from 100 $\mu$Jy beam $^{-1}$(3 $\times$ rms) and increase by factors of 2. S409 is located at the edge of the 3.6-$\mu$m image.}\label{wats}
\end{figure*}

\begin{figure*}
\begin{center}
\begin{tabular}{ccc}
\includegraphics[angle=0, scale=0.9]{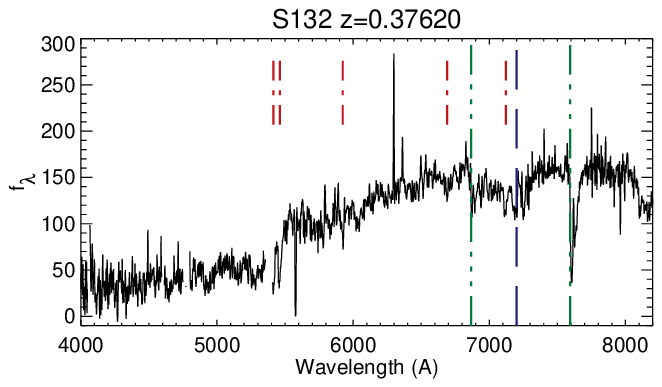}& \includegraphics[angle=0, scale=0.9]{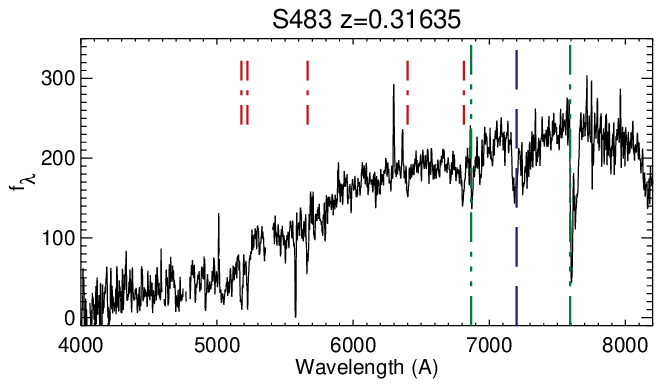} \\
\includegraphics[angle=0, scale=0.9]{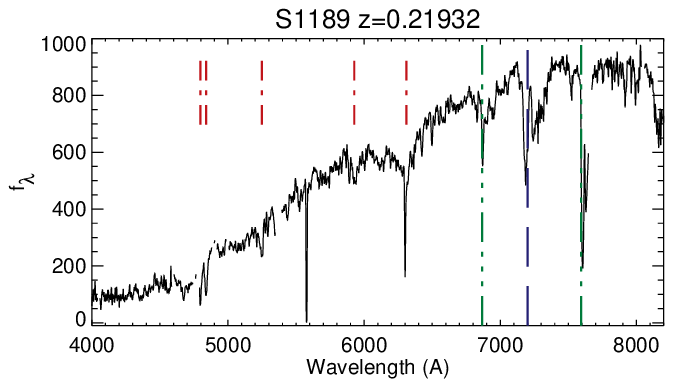} & \includegraphics[angle=0, scale=0.9]{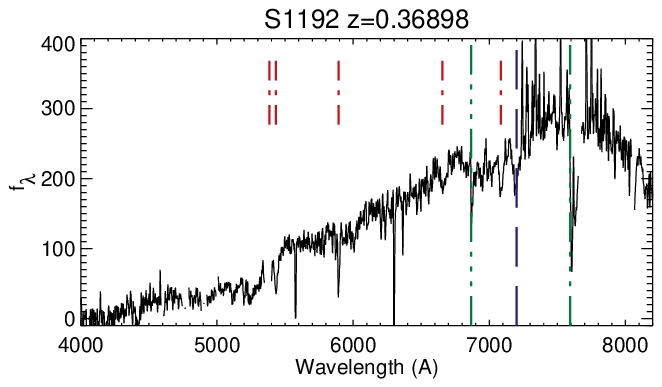}\\
\includegraphics[angle=0, scale=0.9]{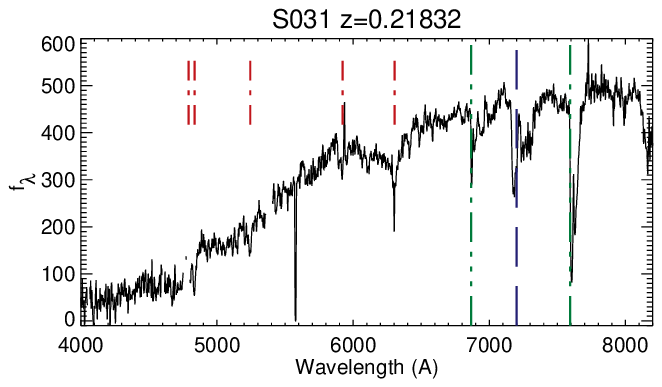}\\
\end{tabular}
\end{center}
\caption{Spectra of the host galaxies of the five WATs for which we have measured redshifts using the AAOmega spectrograph on the AAT. The spectra are typical of early-type galaxies that host luminous radio sources. The red dot-dashed lines indicate the prominent stellar absorption features typical of an early-type galaxy spectrum (Ca H+K, G-band, H-beta and Mg-b) from which the redshift has been derived via template cross correlation. The green dash-dot-dot-dot lines indicate the Fraunhofer A+B atmospheric absorption bands from O$_2$ and the blue long dashed lines indicate the atmospheric water absorption band, neither of which
have been corrected due to the absence of appropriate telluric standards in the redshift survey data.}\label{watspec}
\end{figure*}

We have identified six WATs in ATLAS by visually examining the
greyscale ATLAS images \citep{Norris06,Middelberg08}. Fig. \ref{wats}
shows the ATLAS greyscale radio images of the WATs in the left column,
while images of the WATs superposed on the Digitized Sky Survey (DSS)
red and 3.6-$\mu$m Infrared Array Camera (IRAC) images are shown in
the middle and right columns respectively. The WATs range in redshift
from 0.1469 to 0.3762, and their properties are summarized in Table
\ref{watsinatlas}. The radio luminosities at 1.4 GHz range from
$\sim$2$-$6$\times$10$^{24}$ W Hz$^{-1}$ which places them in the FRI
\citep{Fanaroff74} category. For comparison the median luminosities of
radio sources associated with cD galaxies in rich and poor clusters
studied by \citet{Giacintucci07} are 0.7$\times$10$^{24}$ and
0.2$\times$10$^{24}$ W Hz$^{-1}$ at 1.4 GHz. We have estimated the
absolute R-band magnitudes of our ATLAS sources and find that these
lie close to the transition region in the absolute red-magnitude$-$1.4
GHz radio luminosity plot of \citet{Owen94}. The optical spectra of
the five sources for which we have determined redshifts, of which four
(S132, S483, S1189 and S1192) are new, are presented in
Fig. \ref{watspec}. The redshift of the sixth WAT galaxy, S409, was
determined by Colless et al. (2001).

We have probed for overdensities of galaxies in the vicinity of the
WATs.  In addition to our observations of S1189 mentioned earlier, we
have examined the 2dFGRS \citep{Colless01} spectroscopic survey, as
well as the photometric redshifts of galaxies in the {\it SWIRE} field
by \citet{Rowan-Robinson08}.  The 2dFGRS shows an overdensity of
galaxies associated with S409, which is the nearest WAT in our sample,
while the photometric redshifts indicate overdensities of galaxies
associated with S483 and S1192.

We have examined archival \emph{ROSAT} All-Sky Survey (RASS) data for
X-ray detections, and found no RASS detections towards these WATs.
This implies an upper limit to the X-ray luminosity of potential host
clusters of $\sim$2$-$11$\times$10$^{37}$ W s$^{-1}$ which spans the
upper values typical for clusters of galaxies with known X-ray
emission \citep{Boehringer01}.  This indicates upper limits to the
masses of $\sim$2$-$6$\times$ 10$^{14}$M$_{\odot}$ \citep{Pratt09}.

\begin{table*}
\begin{center}
  \caption{Sample of WATs in ATLAS. Columns 1 and 2 give the ATLAS and
    SWIRE names, Column 3 gives the redshift.  Column 4 lists the
    observed R-band magnitude from SWIRE, except for S132 where we
    have listed the value from superCOSMOS since a value from SWIRE is
    not available, while Column 5 lists the absolute R-band magnitude.
    Columns 6 and 7 list the flux density and luminosity respectively
    at 1.4 GHz. Columns 8 and 9 list the angular and physical
    size. All the WAT redshifts were obtained from our AAT
    observations with the exception of S409 whose redshift was
    determined by \citet{Colless01}.}
\begin{tabular}{lllllrccr}
\hline
ATLAS & SWIRE Counterpart &z &
R$_{\rm obs}$ &R$_{\rm abs}$ & 
Flux$_{1.4}$ & Power$_{1.4}$ & Size$_{ang}$ & Size$_{phys}$\\
& & & (mag) & (mag) & (mJy) & (10$^{24}$W/Hz) & (arcmin) & (kpc)\\
\hline
ELAIS-1 \\
S132 & SWIRE4\_J003236.18-442101.1  & 0.3762 & 18.1  & $-$23.41 & 5.35 & 2.58 & 1.0 & 309\\
S483 & SWIRE4\_J003311.21-435512.3  & 0.3164 &18.28 &$-$22.79 & 6.72 & 2.16 & 1.5 & 413\\
S1189 & SWIRE4\_J003427.54-430222.5 & 0.2193 &17.12 &$-$23.04 & 45.03 & 6.25 & 5.0 & 1053\\
S1192 & SWIRE4\_J003320.68-430203.6 & 0.3690 &18.92 &$-$22.54 & 10.46 & 4.82 & 0.9 & 274\\
CDFS \\
S031 & SWIRE3\_J032639.11-280801.5  & 0.2183 &16.63 &$-$23.52 & 42.29 & 5.81 & 2.7 & 567\\
S409 & SWIRE3\_J033210.74-272635.5  & 0.1469 &16.35 &$-$22.84 & 42.35 & 2.41 & 2.6 & 396\\
\hline
\label{watsinatlas}
\end{tabular}
\end{center}
\end{table*}

\subsection{S132}
The largest angular size of the source from end to end along the axis
of the source is 0.96 arcmin, corresponding to a physical size of
$\sim$309 kpc. The peak of emission to the southwest of the host
galaxy has been determined to be an unrelated source, S131
\citep{Middelberg08}.  There are several galaxies to the west of the
southern tail which are seen more clearly in the 3.6-$\mu$m image, but
at present no redshift information is available for these galaxies. The tails
appear to bend away from this overdensity of galaxies.

\subsection{S483}
\begin{figure}
\begin{center}
\includegraphics[scale=0.25, angle=-90]{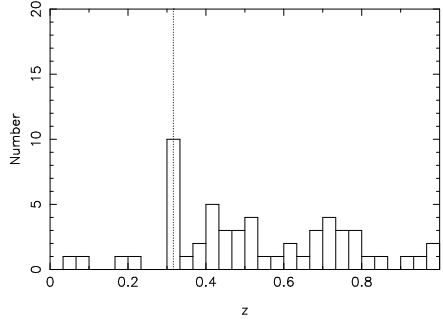}
\end{center}
\caption{Photometric redshift distribution of galaxies within 2 arcmin of S483 ($\sim$550 kpc at z = 0.3164). The data is binned in intervals of $\Delta$z = 0.03. The vertical dotted line indicates the redshift of the host galaxy.}\label{S483photz}
\end{figure}

This WAT, which has an overall linear size of 413 kpc, is highly
asymmetric in the brightness of the two tails, with the peak
brightness in the northern tail being higher by a factor of
$\sim$5. It would be useful to image the source, especially
the southern tail, with higher surface brightness sensitivity to
confirm the present classification. The photometric redshifts of the galaxies
\citep{Rowan-Robinson08} within a radius of 2 arcmin, which
corresponds to $\sim$550 kpc at z=0.3164, show a concentration of
galaxies at about the redshift of S483 (Fig. \ref{S483photz}).

\subsection{S1189}
S1189 is the largest WAT in our sample with an overall linear size of 
$\sim$1053 kpc. Its opening angle, defined by the lines connecting the
regions of highest surface brightness to the optical galaxy is $\sim$70$^\circ$,
which is slightly smaller than for the high-redshift WAT reported by \citet{Blanton01}
which has an opening angle of $\sim$80$^\circ$. Clearly these opening angles
would depend on the resolution of the observations and projection effects. 
\citet{Rudnick77} distinguish between narrow-, intermediate- and
wide-angle tails by requiring that the opening angle be less than $\sim$20$^\circ$
for narrow-angle tailed sources and greater than $\sim$90$^\circ$ for WATs, 
based largely on tailed sources at smaller redshifts than our sources. Although
it would be relevant to examine the effects of resolution and surface brightness
sensitivity as one finds more tailed sources at moderate and high redshifts, the
opening angle of S1189 is close to that of a WAT. Although WATs do tend to be
associated with the dominant galaxy, it could be associated with a bright galaxy
close to the brightest galaxy in a cluster or group \citep[see][]{Rudnick77,
Blanton01}. The associated galaxy of S1189 is the next brightest galaxy, only 
0.75 mag fainter than the cD galaxy. \citet{Rudnick77} also suggested that WATs
tend to have larger sizes than the narrow-angle tailed sources. With a total 
size of over a Mpc, it would be more consistent with the sizes of WATs. 
Considering all the aspects, we presently classify it as a WAT.         
We discuss the results of our AAOmega observations and 
the environment of this source  in Section \ref{S1189}.

\subsection{S1192}
\begin{figure}
\begin{center}
\includegraphics[scale=0.25, angle=-90]{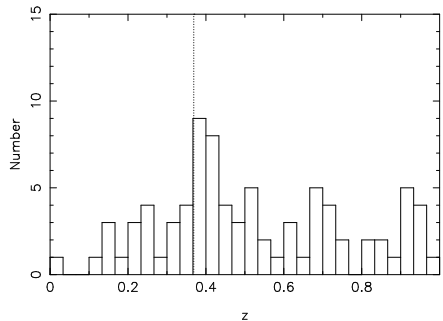}
\end{center}
\caption{Photometric redshift distribution within 2 arcmin of S1192 ($\sim$500 kpc at z = 0.3690). The data is binned in intervals of $\Delta$z = 0.03. The vertical dotted line indicates the redshift of the host galaxy.}\label{S1192photz}
\end{figure}

S1192 is similar to S132 in both shape and extent, but the two tails
in S1192 are more symmetric in brightness.  Both the DSS red and
3.6-$\mu$m images show a number of galaxies forming a filamentary-like
structure along with the host galaxy of the WAT.  The photometric
redshifts \citep{Rowan-Robinson08} within a radius of 2 arcmin, which
corresponds to $\sim$500 kpc at z=0.3690, show a concentration of
galaxies at about the redshift of S1192 (Fig. \ref{S1192photz}).
This overdensity is largely due to the galaxies in the
filamentary-like structure.

\subsection{S031}
Although S031 exhibits distinct gaps of emission between the radio
core and the two tails of emission, the identification process
described by \citet{Norris06} unambiguously classifies these three
components as a triple radio source. The peaks of emission in the
tails are towards the radio core as expected in FRI radio sources.  We
do not have redshift information at present to determine which of the
galaxies seen in Fig. \ref{wats} may be a part of the group or cluster
associated with S031. The gaps of emission between the central
source and the lobes are reminiscent of the large radio galaxy in
Abell 2372 \citep{Owen97, Giacintucci07}, which has been suggested by
Giacintucci et al. (2007) to be due to recurrent radio activity
(see Saikia \& Jamrozy 2009 for a review). Although such a possibility
cannot be ruled out, more detailed spectral and structural information
are required to clarify whether this is indeed the case.  

\subsection{S409}
\begin{figure}
\begin{center}
\includegraphics[angle=-90, scale=0.25]{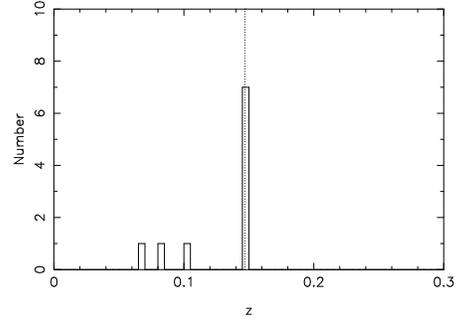}
\end{center}
\caption{2dFGRS spectroscopic redshifts \citep{Colless01} of sources
  within 7 arcmin of S409 ($\sim$1 Mpc at z = 0.1469). The data is
  binned in intervals of $\Delta$z = 0.005. The vertical dotted line
  indicates the redshift of the host galaxy.}\label{S409spectrum_2df}
\end{figure}

S409 is the closest of the WATs in ATLAS with a redshift of 0.1469
\citep{Colless01} and a size of 396 kpc.  It has an interesting radio
structure with the western lobe exhibiting two sharp bends and forming
a long narrow tail of emission. A deep X-ray image would be useful to
understand how the gas distribution may have shaped the unusual radio
structure. The 2dFGRS data \citep{Colless01} within a radius of
$\sim$1 Mpc show a clear excess of galaxies in the same redshift bin
as S409 (Fig. \ref{S409spectrum_2df}). Within a radius of 500 kpc (3.5
arcmin at z = 0.1469) we find 3 galaxies at about the redshift of
S409.

\section{Large-Scale Structure Around S1189}\label{S1189}

There are a total of 309 galaxies with spectroscopic
redshifts within a radius of one degree of S1189, including 94 galaxies
whose redshifts we have measured from our service mode observations
with the AAOmega spectrograph. The other redshifts are obtained from
spectroscopic observations of ATLAS sources (Section
\ref{specdata}). The redshifts of these 94 new galaxies are listed in
Appendix A.

\subsection{Redshift distribution}
\begin{figure*}
\begin{center}
\includegraphics[angle=-90, scale=0.5]{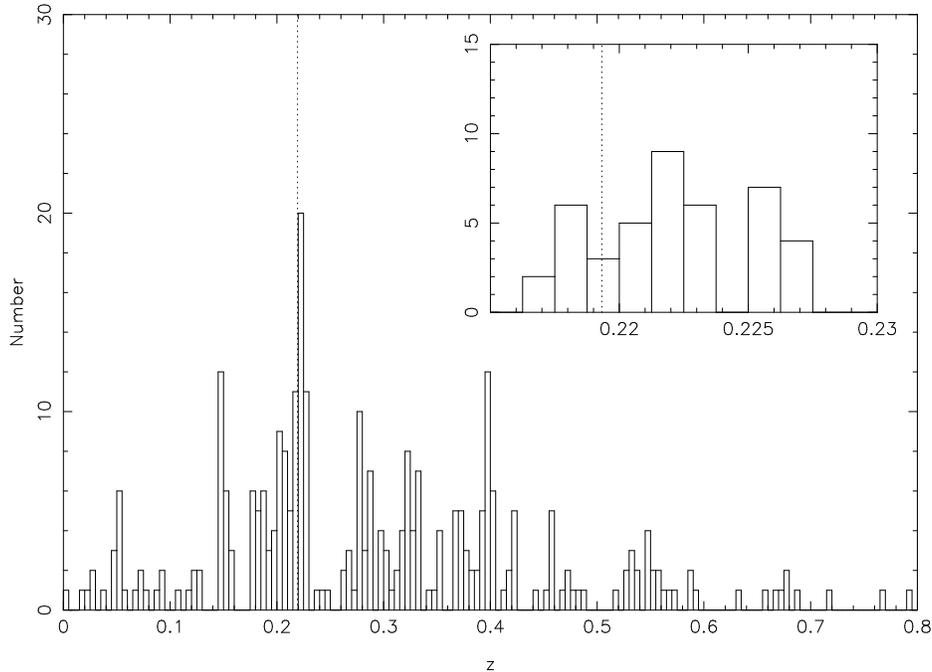}
\end{center}
\caption{Histogram of redshifts of 299 galaxies located within a one degree radius centred on the WAT, S1189, and with redshifts less than or equal to 0.8. The redshift bin size is 0.005 which corresponds to 1500 km s$^{-1}$. The inset shows the redshift distribution of the galaxies at the peak and two adjacent bins (0.215 $\leq$ z $<$ 0.23). The redshift bin size is 0.00125 which corresponds to 375 km s$^{-1}$. The vertical dotted lines in both histograms are at z=0.2193, 
the redshift of the WAT source.}\label{z_hist}
\end{figure*}

The redshifts of the 309 galaxies within a radius of one degree
($\sim$12.6 Mpc at z $\sim$ 0.22) from the WAT source extend to
$\sim$1.95.  The distribution for the subset of 299 galaxies with
z$\leq$0.8 is shown in Fig. \ref{z_hist}. The data are binned in
intervals of $\Delta$z = 0.005 which corresponds to 1500 km s$^{-1}$.
There is a clear excess of galaxies at the redshift of the WAT source,
with a distinct peak at the redshift bin 0.22 $\leq$ z $<$ 0.225. 20
galaxies lie in the peak-redshift bin, and a further 22 galaxies lie
in the two neighbouring bins resulting in 42 galaxies over three
redshift bins, the concentration being significant at $\sim$7$\sigma$.
Properties of the galaxies in the peak histogram bin and the two
adjacent bins, which includes the putative cD galaxy at a redshift of
0.2204, are listed in Table \ref{peak}.  
The total spread in velocity of the 42 galaxies is $\sim$4500 km
s$^{-1}$, and the velocity dispersion is $\sim$870 km s$^{-1}$.  
This is similar to the spread for typical rich clusters in
the local Universe undergoing mergers such as A3667 and A3376 which
both show \textbf{radio} relic emission and have a velocity spread of $\sim$4200 km
s$^{-1}$ \citep{mjh08, mjh10, Owers09}.  The redshift distribution of
the 42 galaxies is shown in greater detail as an inset in
Fig. \ref{z_hist}. The distribution is not a smooth Gaussian and shows
sub-structure, consistent with dynamic, merging systems.

\begin{table*}
\begin{center}
\caption{Optical and infrared properties of the putative cluster members. Column 1 gives the SWIRE identification.  Column 2 lists the R-band magnitude from SWIRE while Column 3 lists the redshifts. Column 4 lists the radio flux density at 1.4 GHz while Column 5 provides comments.}\label{peak}
\begin{tabular}{ccccl}
\\
\hline
\multirow {2}{*}{SWIRE ID} & \multirow {2}{*}{R mag}& \multirow {2}{*}{z} & Radio flux& \multirow {2}{*}{comment}\\
&&&(mJy)&\\
\hline
SWIRE3\_J003236.91-432040.8 &  19.44 &  0.2169 &   &  \\
SWIRE4\_J003411.57-425952.0 &  18.92 &  0.2171 &  0.53 &  \\
SWIRE4\_J003107.43-434037.5 &   &  0.2176 &  0.24 &  \\
SWIRE4\_J003559.43-430324.8 &  18.74 &  0.2180 &  0.21 &  \\
SWIRE4\_J003109.85-435010.9 &   &  0.2181 &  3.93 &  \\
SWIRE4\_J003512.31-425437.5 &   &  0.2181 &  14.79 &  \\
SWIRE3\_J003203.05-434121.6 &  18.07 &  0.2184 &   &  \\
SWIRE3\_J003500.92-430309.5 &  20.06 &  0.2187 &   &  \\
SWIRE3\_J003355.92-424153.9 &   &  0.2190 &   &  \\
SWIRE4\_J003123.87-430940.5 &   &  0.2191 &  0.34 &  \\
SWIRE4\_J003427.54-430222.5 &  17.12 &  0.2193 &  45.03 & WAT \\
SWIRE3\_J003422.08-430623.7 &  19.53 &  0.2201 &   &  \\
SWIRE4\_J003432.80-424555.1 &   &  0.2202 &  1.08 &  \\
SWIRE4\_J003748.72-430211.9 &  18.65 &  0.2203 &  0.23 &  \\
SWIRE4\_J003525.13-432941.4 &  18.93 &  0.2204 &  0.15 &  \\
SWIRE3\_J003419.26-430334.0 &  16.37 &  0.2204 &   & cD galaxy \\
SWIRE4\_J003713.54-431342.8 &  17.53 &  0.2214 &  2.39 &  \\
SWIRE3\_J003339.84-430908.8 &  18.07 &  0.2215 &   &  \\
SWIRE3\_J003711.92-430711.4 &  18.32 &  0.2215 &   &  \\
SWIRE3\_J003415.87-430840.9 &  18.97 &  0.2220 &   &  \\
SWIRE4\_J003714.11-430833.3 &  17.59 &  0.2221 &  0.53 &  \\
SWIRE3\_J003526.70-430418.7 &  18.45 &  0.2222 &   &  \\
SWIRE3\_J003503.98-425710.2 &   &  0.2222 &   &  \\
SWIRE4\_J003645.81-432016.0 &  17.73 &  0.2222 &  0.49 &  \\
SWIRE3\_J003443.66-424544.6 &   &  0.2225 &   &  \\
SWIRE4\_J003344.79-431627.8 &  17.87 &  0.2228 &  0.91 &  \\
SWIRE3\_J003707.12-430302.7 &  19.18 &  0.2229 &   &  \\
SWIRE4\_J003242.01-432630.6 &  18.78 &  0.2230 &  0.32 &  \\
SWIRE4\_J003326.18-434051.0 &  18.76 &  0.2232 &  0.27 &  \\
SWIRE3\_J003229.91-425457.7 &   &  0.2233 &   &  \\
SWIRE3\_J003242.01-432630.5 &  18.78 &  0.2233 &   &  \\
SWIRE4\_J003721.05-434240.0 &  17.40 &  0.2251 &  1.33 &  \\
SWIRE4\_J003306.30-431029.8 &  17.71 &  0.2252 &  12.33 & double radio  \\
SWIRE4\_J003609.95-435002.2 &  19.84 &  0.2252 &  0.32 &  \\
SWIRE3\_J003322.00-430419.5 &  20.12 &  0.2253 &   &  \\
SWIRE4\_J003604.09-435802.3 &   &  0.2255 &  0.18 &  \\
SWIRE4\_J003340.23-432542.2 &  18.24 &  0.2258 &  0.34 &  \\
SWIRE4\_J003640.42-430000.1 &  17.32 &  0.2263 &  0.75 &  \\
SWIRE4\_J003734.09-433339.3 &  17.92 &  0.2263 &  1.20 &  \\
SWIRE3\_J003659.30-431824.1 &  18.20 &  0.2263 &   &  \\
SWIRE4\_J003502.52-432410.9 &  18.02 &  0.2265 &  0.19 &  \\
SWIRE3\_J003300.09-432819.9 &  18.26 &  0.2267 &   &  \\

\hline
\end{tabular}
\end{center}
\end{table*}

\subsection{Spatial Distribution}
\begin{figure*}
\begin{center}
\includegraphics[angle=-90, scale=0.34]{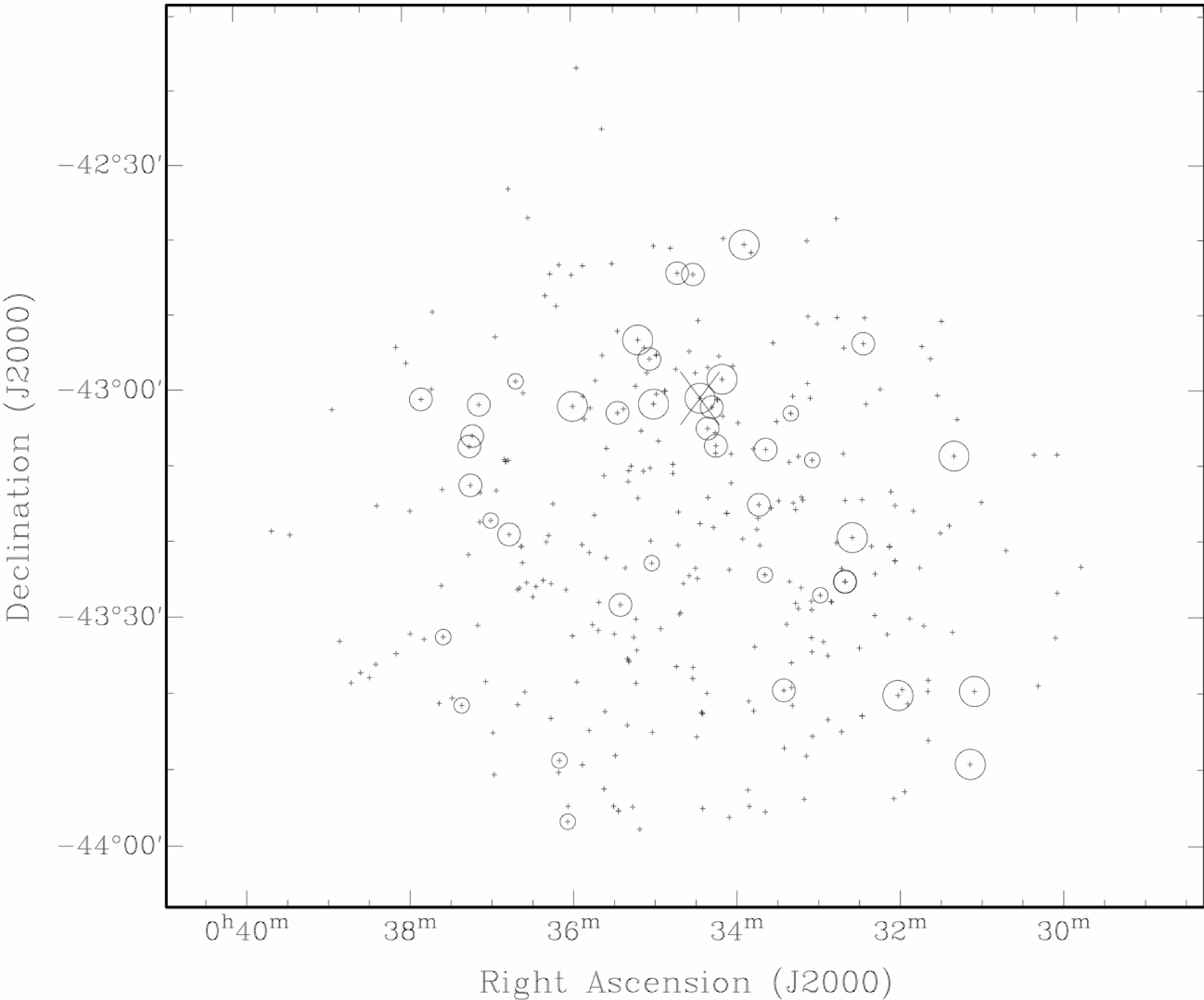}
\end{center}
\caption{Spatial distribution of all galaxies in the field surrounding S1189 that have spectroscopic redshifts. 
The circled sources are at 0.215 $\leq$ z $<$ 0.23. The largest circles show the sources that are at 0.215 $\leq$ z $<$ 0.22. The medium circles show the sources that are at 0.22 $\leq$ z $<$ 0.225 and the smallest circles show the sources that are at 0.225 $\leq$ z $<$ 0.23. The location of the WAT is indicated by the large ``X''.}\label{spatialdist}
\end{figure*}

In Fig. \ref{spatialdist} we plot the positions of the 42 galaxies listed in Table
\ref{peak}, with the galaxies in the three
redshift bins (0.215 $\leq$ z $<$ 0.22; 0.22 $\leq$ z $<$ 0.225; 0.225
$\leq$ z $<$ 0.23) indicated by circles of varying size. Despite
considerable overlap, there is a suggestion of a velocity gradient
with the galaxies in the lowest redshift bin (largest circles) extending
towards the south-west and those in the highest redshift bin (smallest
circles) extending towards the south-east.  Although galaxy redshifts
were measured within a radius of $\sim$12 Mpc from the WAT, $\sim$60
per cent of the galaxies listed in Table \ref{peak} are within 6 Mpc
of the WAT (30 arcmin). We also note that the larger number of sources
in the southern part of Fig. \ref{spatialdist} is due to the uneven
coverage of the one-degree-radius field surrounding S1189.

\subsection{cD Galaxy}
The bright galaxy, SWIRE3\_J003419.26-430334.0, located southwest of
the WAT source, has a redshift of 0.2204, implying a velocity
difference between the two galaxies of $\sim$320 km s$^{-1}$. Their
projected separation is $\sim$2 arcmin, corresponding to $\sim$420
kpc at a redshift of 0.22.  SWIRE3\_J003419.26-430334.0 is the
brightest galaxy in the cluster and has a diffuse envelope, therefore
we classify it as a possible cD galaxy. There is a marginal detection
of associated radio emission with a flux density of $\sim$140
$\mu$Jy at 1.4 GHz which corresponds to a radio luminosity of 1.96
$\times$ 10$^{22}$ W Hz$^{-1}$. Centrally dominant cD galaxies are
usually giant ellipticals residing in the centres of clusters of
galaxies. These are much larger and brighter than other galaxies in
the cluster and are often surrounded by a diffuse envelope
\citep{Matthews64}.  Their large size is usually attributed to mergers
and galaxy cannibalism \citep[e.g.][]{DeLucia07}.

\subsection{Extended radio sources in the vicinity of the WAT}
In addition to double-lobed radio sources, radio haloes, relics and
core haloes or mini haloes may also be associated with clusters of
galaxies. Core-haloes are usually less than $\sim$500 kpc in extent
and associated with the dominant galaxy in cooling core
clusters. Haloes and relics are not associated with any particular
galaxy, and are often larger in size.  Radio haloes are usually
projected towards the cluster centre, while relics are seen towards
the periphery \citep[e.g.][]{Giovannini04}.  There are $\sim$30
radio haloes in nearby (z$<$0.4) clusters of galaxies
\citep[e.g.][]{Giovannini09}, and there are $\sim$30 clusters of
galaxies with at least one radio relic \citep{Giovannini04}.
While models for haloes range from re-acceleration of particles by 
turbulence to production of relativistic electrons by hadronic collisons, 
relics are believed to arise due to cluster mergers and/or matter accretion 
\citep{Sarazin99, Ryu03, Pfrommer06, Giacintucci08, mjh08,Brown09}.

Recent work suggests that halos are found in massive,
unrelaxed clusters, with the radio and X-ray luminosity being strongly
correlated, consistent with the re-acceleration scenario 
\citep{Brunetti07, Venturi08, Cassano09}. However, the present studies
have been based on X-ray selected clusters of galaxies, and possible
biases arising from it should be borne in mind. For example, the limited 
sensitivity of the radio observations would make it easier to detect halos 
in only the more X-ray luminous clusters of galaxies. 

Radio relics on the
other hand are believed to arise due to mergers accompanied by shocks
and/or matter accretion (e.g. Bagchi et al. 2006, and references therein).
These shocks are capable of accelerating particles to high energies, giving rise 
to the observed synchrotron radio emission. \citet{Harris80} and \citet{Tribble93}
were amongst the early ones to suggest and explore the possibility of
acceleration of particles due to shock fronts on a large scale caused by
mergers. These ideas were expanded upon by \citet{Ensslin98}, \citet{Roettiger99}
\citet{Ensslin01} and \citet{Ricker01}, producing more sophisticated models.

\begin{figure*}
\begin{center}
\includegraphics[angle=-90, scale=0.7]{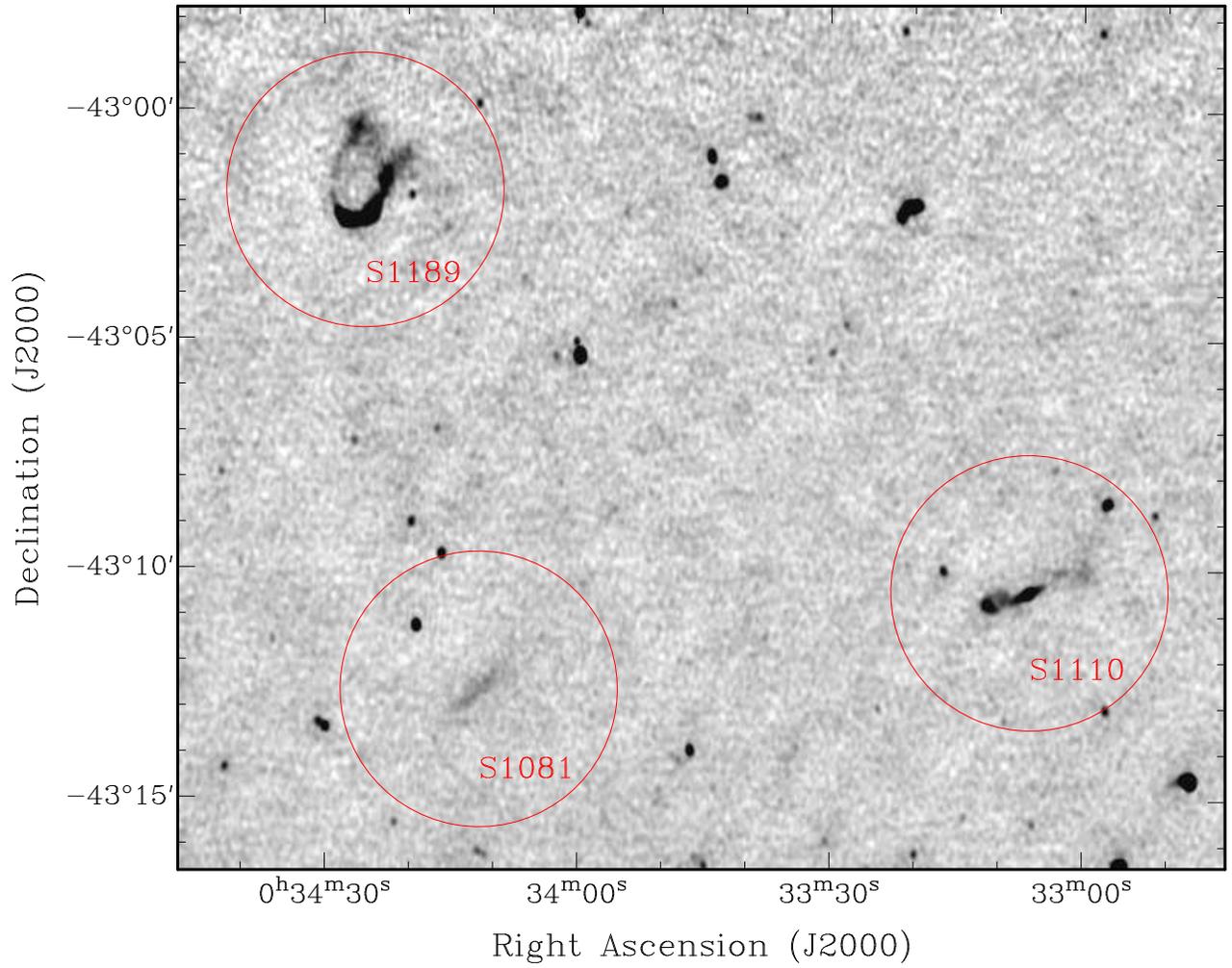}
\end{center}
\caption{1.4\,GHz radio image showing the WAT (S1189), the
  double-lobed radio galaxy (S1110) and the radio relic (S1081). The
  WAT S1192 , at z = 0.3690, is also seen in the
  image.\label{disturbed}}
\end{figure*}

\begin{figure*}
\begin{center}
\includegraphics[angle=-90, scale=0.3]{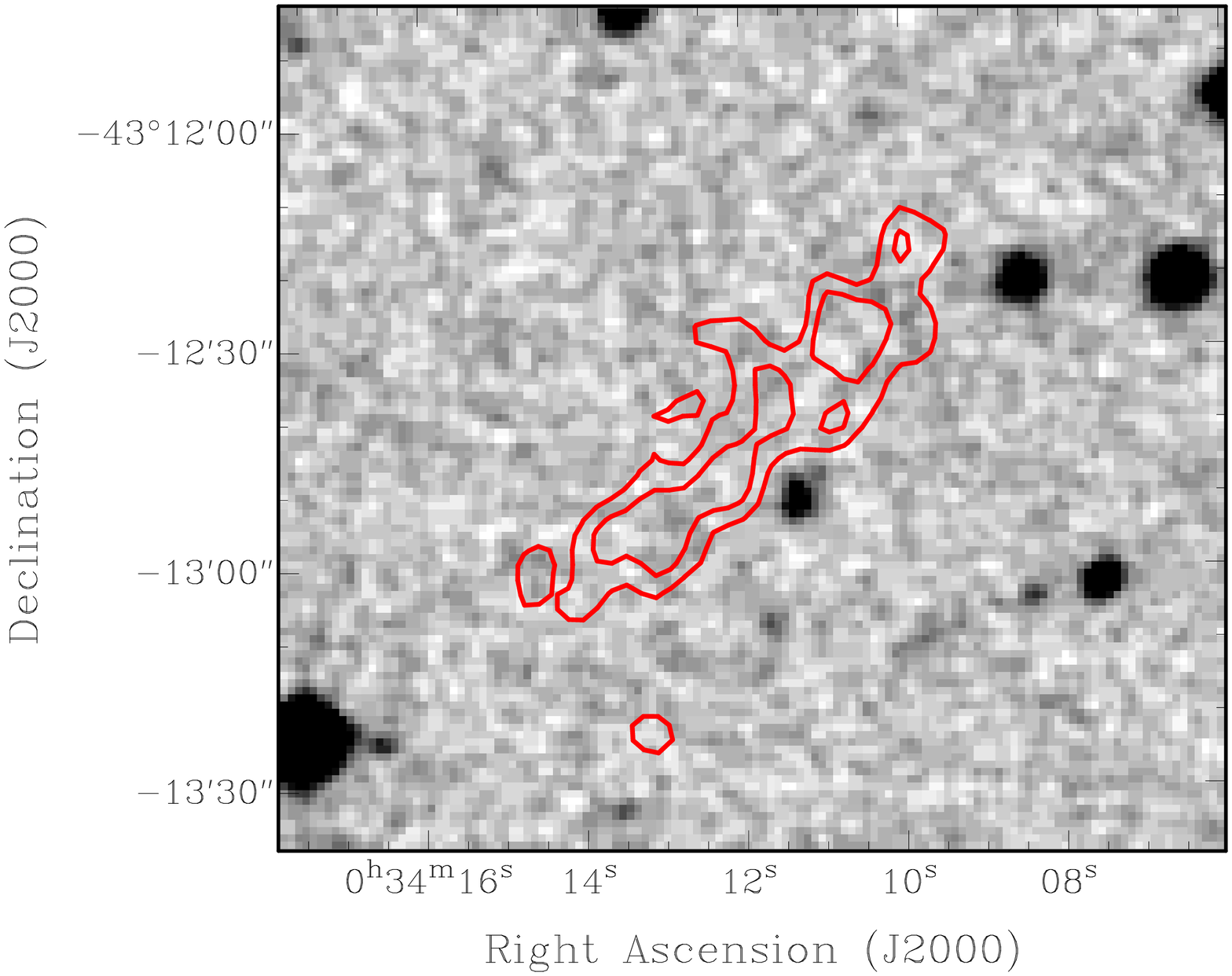}
\includegraphics[angle=-90, scale=0.3]{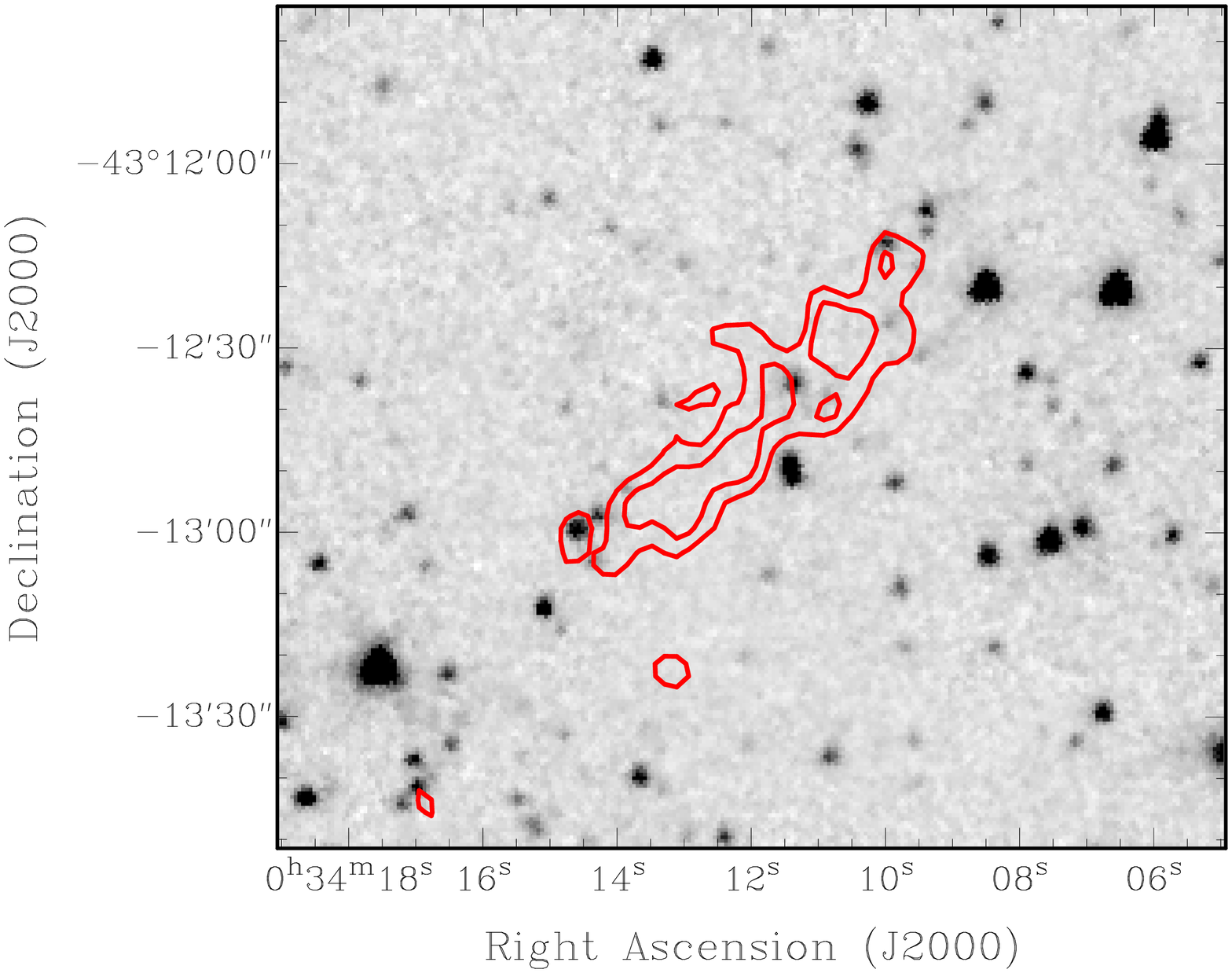}
\end{center}
\caption{ 1.4\,GHz radio contours of the relic (S1081) overlaid on the
DSS red optical image (left panel) and the 3.6 $\mu$m IRAC image (right panel)
The contours start at 100 $\mu$Jy beam$^{-1}$ and increase by factors of $\sqrt{2}$.} 
\label{relic}
\end{figure*}

The ATLAS radio image at 1.4 GHz (Fig. \ref{disturbed}) shows two more
extended sources within $\sim$20 arcmin of the WAT source,
one of which (S1081) appears to be a radio relic \citep{Middelberg08}, 
while the other (S1110) is an FRI radio galaxy. Superpositions of
the radio image of the relic on an optical DSS red image as well as an
infrared 3.6$-\mu$m image are shown in  Fig.~\ref{relic}.
While no optical object is visible within the radio contours, there is an infrared 
object towards the central region of the source. This object has been classified 
as an Sbc galaxy (optical template type 5) using a
total of 6 photometric bands by Rowan-Robinson et al. (2008). Its photometric 
redshift has been estimated to be 1.18. Given the properties of the object, it 
is likely to be unrelated.
The radio properties of the WAT and these two
sources are summarised in Table \ref{disturbedtable}. At a redshift of
0.22, the relic would have a physical size of $\sim$274 kpc and a
luminosity of \textbf{3.3}$\times$10$^{23}$ W Hz$^{-1}$, which would make it
similar to the relics found in the periphery of clusters of galaxies
in the local Universe \citep{Ferrari08}. The relic is at a projected distance of
$\sim$2 Mpc from the cD galaxy. Typically relics have been observed at
distances of about a Mpc from the cluster centre, although some
systems are known to have relics up to distances of $\sim$4 Mpc
\citep[e.g.][]{Giovannini04}. Some of the known examples of relics which lie at 
distances beyond $\sim$2 Mpc from the nearest cluster core, such
as B0917+75 (Harris et al. 1993; Johnston-Hollitt 2003), are typically associated 
with structure larger than a single cluster. In the case of B0917+75 it is the 
Rood 27 cluster group. This is similar to our situation. 
It is also relevant to note that the minor axis
of the relic does not point towards either the WAT source or the cD
galaxy, suggesting sub-structure in this large-scale structure.  
Simulations of shock generation during
hierarchical mass assembly suggest relics can be produced over 8 Mpc
from the cluster centre \citep{Miniati00, Pfrommer06, Pfrommer08,
Hoeft08, Vazza09}. These aspects along with its radio structure and
lack of an obvious optical identification make it very likely to be a radio
relic. One could enquire whether this object might be a dying radio galaxy.
The non-detection of an early-type galaxy associated with it suggests that
this is unlikely to be the case.  There
are very few relics known beyond a redshift of $\sim$0.2
\citep[e.g.][]{Giovannini04}, which makes this finding a significant one.

The other interesting source in the field is the FRI radio source S1110.
The radio emission from S1110 is symmetric
within $\sim$80 kpc from the host galaxy, SWIRE4\_J003306.30-431029.8,
reminiscent of the large-scale jets in FRI radio sources. However, the
extended lobes are highly asymmetric, the peak brightness in the outer
extremities differing by a factor of $\sim$4. This may be due to
density asymmetries on opposite sides of the source.

\begin{table*}
\begin{center}
\caption{\label{disturbedtable}Radio properties of S1189 and extended radio sources in its vicinity. The size of the 
WAT (S1189) was measured from the outer edge of one lobe to the core and out to the outer edge of the other lobe. 
The relic is assumed to be at a redshift of 0.22.}
\begin{tabular}{lcccccccc}
\\
\hline
& \multirow {2}{*}{ATLAS ID} & RA & Dec  & \multirow {2}{*}{Redshift} & S$_{1.4}$ & Power$_{1.4}$ & size$_{ang}$ & size$_{phy}$ \\
& &(J2000) & (J2000)& & (mJy) & (10 $^{24}$ W Hz$^{-1}$) & (arcmin) & (kpc)\\
\hline
WAT & S1189& 00 34 27.6 & -43 02 22.5 & 0.2193 & 45.03 & 6.25 & 5.0 & 1053\\
Double radio & S1110 & 00 33 06.3 & -43 10 29.8  & 0.2252& 12.33 & 1.82 & 2.6 & 559\\
Relic & S1081 & 00 34 11.7 & -43 12 39.4 & (0.22)  & 2.35&  (0.33) & 1.25 & (274) \\
\hline
\end{tabular}
\end{center}
\end{table*}

\section{Implications for Deep Wide Radio Surveys and ATLAS}

In this paper we have reported the detection of six WATs from a sample
of 2004 radio sources. Extrapolating this to future deep wide surveys,
we might expect to detect about 200,000 WATs from the catalogue of 70
million radio sources that will be generated by the ASKAP-EMU
(Australia SKA Pathfinder - Evolutionary Map of the Universe) project
\citep{Norris09}. Since each of these WATs is likely to be associated
with a cluster, such surveys will be powerful tools for detecting
clusters and exploring their properties, particularly since the radio
luminosity of WATs makes them detectable and capable of being studied
up to high redshifts.

Such surveys are therefore likely to contribute significantly to areas
such as the formation and evolution of clusters, the formation of
massive ellipticals, and the relationship between giant
ellipticals and supermassive black holes, or SMBHs
\citep[e.g.][]{Blanton03, Chiaberge09}. Furthermore, while optical and
X-ray surveys tend to select the optically-rich or most X-ray-luminous
clusters of galaxies at moderate and high redshifts, sensitive radio
observations could help identify clusters with a wide range of optical
and X-ray properties.

However, WATs are characterized by diffuse lobes of emission extending
to hundreds of kpc, and two effects potentially make such structures
difficult to observe at high redshift.

First, the radiating electrons of a synchrotron source lose energy by
inverse-Compton (iC) scattering of the cosmic microwave background
radiation (CMBR), whose energy density increases as (1+z)$^4$. This
effect is supported by evidence that the X-ray emission from the lobes
of large radio galaxies is due to iC scattering of the radiating
electrons with the CMBR, which has been used to make an independent
estimate of the magnetic field strength of the radio lobes
\citep[e.g.][]{Croston04, Croston05, Konar09}. Furthermore,
\citet{Konar04} have found that the bridge emission in giant radio
sources is less prominent at higher redshifts, which they interpret as
being caused by iC scattering with the CMBR.

Loss of electron energy by iC scattering from the CMBR overtakes
synchrotron cooling at a redshift $z \sim 0.556 \sqrt{B} -1 $ where B
is the synchrotron magnetic flux density in $\mu$Gauss
\citep{Schwartz06}. So, for a constant B, one might expect synchrotron
emission to fall sharply above that redshift.

However, if a low-luminosity radio source is modelled as two cones of
expanding plasma on either side of the central SMBH,
then the magnetic field would be expected to fall as the square of the
distance r from the SMBH, resulting in a transition radius r$_{crit}$ at
which the dominant electron cooling mechanism switches from synchrotron
to iC, where $r_{crit} \propto (1+z)^{-1}$. Thus, rather than synchrotron
emission falling sharply above some redshift, the size of the
synchrotron-emitting region shrinks linearly with redshift.

We conclude that, while iC cooling reduces the apparent size of the
emitting region, it does not impose a fundamental redshift limit above
which WATs will be invisible.

Second, high-redshift galaxies are subject to cosmological surface
brightness dimming, \citep[e.g.][and references
therein]{Lanzetta02} which causes the observed
surface brightness per unit frequency interval of a resolved source to decrease as
(1+z)$^3$. Thus, nearby radio galaxies are detectable to much lower
intrinsic surface brightness thresholds than high-redshift sources.

While both these effects are going to present challenges to the
identification of WATs at high redshifts, they accentuate the normal
challenges of resolution and sensitivity, rather than presenting
fundamental limits of observability. In their search for FRI radio
sources in the redshift range 1$<$z$<$2 using the Faint Images of the
Radio Sky at Twenty-Centimeters (FIRST) radio survey, Chiaberge et
al. (2009) find that most of the sources are compact.  Blanton et
al. (2003) have identified a WAT galaxy at z=0.96, while
\citet{Saikia93} and \citet{Saikia87} explored the possibility that
B1222+216 (4C21.35) and B2\,1419+315 might be WAT quasars at redshifts
of 0.435 and 1.547 respectively.

To explore and understand these aspects will require more detailed
modelling and significantly deeper large-scale radio surveys, which is
the primary goal of ASKAP-EMU.

\section{Conclusions}
We have identified a sample of six Wide-Angle Tail (WAT) radio
sources. We present new spectroscopic redshifts for four of these
sources, and find that these WATs lie in the redshift range
0.1469$-$0.3762. We have examined the fields using both spectroscopic
and photometric redshifts of galaxies in the vicinity of the WATs and
find evidence of an overdensity of galaxies in four of these WATs.

From a more detailed study of the field around S1189 we find an
overdensity of galaxies which is spread over $\sim$12 Mpc and has a
velocity spread of $\sim$4500 km s$^{-1}$, and a velocity dispersion
of $\sim$870 km s$^{-1}$.  This large-scale structure
hosts a putative cD galaxy with, at best, weak radio emission, a radio
relic which has a size of $\sim$274 kpc, and an asymmetric FRI radio
galaxy with an extent of $\sim$559 kpc. The peak brightness at the
extremities of the outer lobes of the FRI source differ by a factor of
$\sim$4, possibly due to differences in the environment on opposite
sides.  The minor axis of the relic is not directed towards either the
host galaxy of the WAT or the putative cD galaxy.  This large-scale
structure may represent an unrelaxed system with different
sub-structures interacting or merging with one another.  Therefore,
deep X-ray observations of the field would be very valuable to further
understand this interesting large-scale structure.

WATs are known to occur in clusters of galaxies, and could in
principle be useful tracers of clusters at moderate and high
redshifts. IC cooling of electrons by interaction with CMBR increases
rapidly with z. However, this does not imply a sharp drop in the
number of WATs at high z.  Deep and wide-field surveys, such as the
Evolutionary Map of the Universe (EMU) \citep{Norris09}, should
provide additional information and insights on the range of structures
at moderate and high redshifts. We expect these to be invaluable
probes of large-scale structure.

\section*{Acknowledgements}
We thank Emil Lenc and Jamie Stevens for their help, the ATLAS
team and Mark Birkinshaw for many fruitful discussions, and an anonymous referee whose
comments helped improve the manuscript. MYM acknowledges the support of an
Australian Postgraduate Award as well as Postgraduate Scholarships
from AAO and ATNF. We thank the staff at AAO and ATCA for making these
observations possible. The ATCA is part of the Australia Telescope,
which is funded by the Commonwealth of Australia for operation as a
National Facility managed by CSIRO. This research has also made use of
NASA's Astrophysics Data System.

\label{lastpage}
\appendix
\section[]{New Redshifts}

\begin{table*}
\begin{center}
\caption{New redshifts of galaxies near S1189. }\label{newz}
\begin{tabular}{lllll}
\\
\hline
SWIRE ID & z & & SWIRE ID & z\\
\hline
SWIRE3\_J003134.02-425148.8	&	0.21061	&	&	SWIRE3\_J003443.66-424544.6	&	0.22247	\\
SWIRE3\_J003147.97-432431.8	&	0.44433	&	&	SWIRE3\_J003445.06-425832.2	&	0.42247	\\
SWIRE3\_J003152.98-431701.7	&	0.24509	&	&	SWIRE3\_J003446.92-431108.0	&	0.32148	\\
SWIRE3\_J003203.05-434121.6	&	0.21837	&	&	SWIRE3\_J003446.92-431221.4	&	0.02507	\\
SWIRE3\_J003205.98-432339.9	&	0.39604	&	&	SWIRE3\_J003448.84-424223.5	&	0.38740	\\
SWIRE3\_J003209.95-432147.1	&	0.20451	&	&	SWIRE3\_J003452.68-430124.9	&	0.18383	\\
SWIRE3\_J003223.48-432147.1	&	0.27936	&	&	SWIRE3\_J003455.92-433249.9	&	0.18833	\\
SWIRE3\_J003229.13-434406.2	&	0.35209	&	&	SWIRE3\_J003458.93-430150.5	&	0.31710	\\
SWIRE3\_J003229.91-425457.7	&	0.22330	&	&	SWIRE3\_J003458.95-425637.6	&	0.32933	\\
SWIRE3\_J003236.91-432040.8	&	0.21686	&	&	SWIRE3\_J003459.03-425642.3	&	0.33043	\\
SWIRE3\_J003242.01-432630.5	&	0.22334	&	&	SWIRE3\_J003500.92-430309.5	&	0.21871	\\
SWIRE3\_J003243.83-430936.9	&	0.20711	&	&	SWIRE3\_J003501.04-424205.0	&	0.41316	\\
SWIRE3\_J003243.91-425533.9	&	0.14944	&	&	SWIRE3\_J003503.98-425710.2	&	0.22218	\\
SWIRE3\_J003248.95-425132.5	&	0.21149	&	&	SWIRE3\_J003506.23-425900.7	&	0.12141	\\
SWIRE3\_J003249.79-423818.9	&	0.30177	&	&	SWIRE3\_J003509.89-430642.4	&	0.20673	\\
SWIRE3\_J003251.92-432910.4	&	0.28767	&	&	SWIRE3\_J003513.81-430046.2	&	0.32167	\\
SWIRE3\_J003300.09-432819.9	&	0.22674	&	&	SWIRE3\_J003519.09-431158.8	&	0.17832	\\
SWIRE3\_J003308.14-430217.3	&	0.18372	&	&	SWIRE3\_J003526.70-430418.7	&	0.22216	\\
SWIRE3\_J003309.95-430020.2	&	0.37209	&	&	SWIRE3\_J003526.75-435641.2	&	0.32339	\\
SWIRE3\_J003310.94-424121.5	&	1.24725	&	&	SWIRE3\_J003527.25-425327.0	&	0.04506	\\
SWIRE3\_J003312.88-431547.4	&	0.27951	&	&	SWIRE3\_J003530.92-424426.3	&	0.53011	\\
SWIRE3\_J003313.91-432722.2	&	0.33233	&	&	SWIRE3\_J003535.18-430900.6	&	0.32241	\\
SWIRE3\_J003317.94-432925.6	&	0.19152	&	&	SWIRE3\_J003537.70-422625.0	&	0.03611	\\
SWIRE3\_J003322.00-430419.5	&	0.22525	&	&	SWIRE3\_J003538.08-425640.0	&	0.26549	\\
SWIRE3\_J003322.79-431047.0	&	0.07282	&	&	SWIRE3\_J003542.74-425959.5	&	0.05296	\\
SWIRE3\_J003335.04-425458.6	&	0.21277	&	&	SWIRE3\_J003551.83-424442.1	&	0.07070	\\
SWIRE3\_J003339.84-430908.8	&	0.22149	&	&	SWIRE3\_J003552.04-430205.0	&	0.18486	\\
SWIRE3\_J003343.91-432149.5	&	0.40215	&	&	SWIRE3\_J003552.98-432142.4	&	0.39342	\\
SWIRE3\_J003348.84-430904.4	&	0.20033	&	&	SWIRE3\_J003556.02-421810.9	&	0.24246	\\
SWIRE3\_J003351.10-424258.2	&	0.26489	&	&	SWIRE3\_J003556.95-433947.0	&	0.42403	\\
SWIRE3\_J003355.92-424153.9	&	0.21902	&	&	SWIRE3\_J003600.06-424555.6	&	0.54820	\\
SWIRE3\_J003400.08-430537.8	&	0.20725	&	&	SWIRE3\_J003609.00-424433.8	&	0.18637	\\
SWIRE3\_J003403.95-425805.9	&	0.32737	&	&	SWIRE3\_J003611.10-425004.1	&	0.33055	\\
SWIRE3\_J003404.81-431335.8	&	0.18986	&	&	SWIRE3\_J003619.16-424839.2	&	0.20139	\\
SWIRE3\_J003404.90-430945.6	&	0.27894	&	&	SWIRE3\_J003630.77-423814.6	&	0.05462	\\
SWIRE3\_J003408.19-431736.3	&	0.14801	&	&	SWIRE3\_J003636.80-432152.6	&	0.15536	\\
SWIRE3\_J003410.81-424105.3	&	0.27872	&	&	SWIRE3\_J003645.06-423419.9	&	0.32288	\\
SWIRE3\_J003410.96-430444.7	&	0.42139	&	&	SWIRE3\_J003645.88-431028.3	&	0.30023	\\
SWIRE3\_J003413.83-425647.6	&	0.32816	&	&	SWIRE3\_J003647.94-431037.1	&	0.29938	\\
SWIRE3\_J003415.22-430234.2	&	0.18821	&	&	SWIRE3\_J003655.07-425404.1	&	0.27827	\\
SWIRE3\_J003415.87-430840.9	&	0.22201	&	&	SWIRE3\_J003659.30-431824.1	&	0.22634	\\
SWIRE3\_J003419.26-430334.0	&	0.22040	&	&	SWIRE4\_J003706.38-431442.3	&	0.66842	\\
SWIRE3\_J003421.99-425817.0	&	0.41943	&	&	SWIRE3\_J003706.70-431836.0	&	0.01959	\\
SWIRE3\_J003422.08-430623.7	&	0.22014	&	&	SWIRE3\_J003707.12-430302.7	&	0.22293	\\
SWIRE3\_J003426.01-434349.1	&	0.20469	&	&	SWIRE3\_J003711.92-430711.4	&	0.22153	\\
SWIRE3\_J003428.82-425203.7	&	0.12161	&	&	SWIRE3\_J003728.02-434143.2	&	0.20727	\\
SWIRE3\_J003430.97-425901.6	&	0.14762	&	&	SWIRE3\_J003748.09-433353.8	&	0.30986	\\

\hline
\end{tabular}
\end{center}
\end{table*}


\begin{thebibliography}{}
\bibitem[\protect\citeauthoryear{Bagchi et al.}{2006}]{Bagchi06}
        Bagchi J., Durret F., Neto G.~B.~L., Paul S., 2006, Sci, 314, 791 
\bibitem[\protect\citeauthoryear{Blanton et al.}{2000}]{Blanton00} 
        Blanton E.~L., Gregg M.~D., Helfand D.~J., Becker R.~H., White R.~L., 2000, ApJ, 531, 118 
\bibitem[\protect\citeauthoryear{Blanton et al.}{2001}]{Blanton01} 
        Blanton E.~L., Gregg M.~D., Helfand D.~J., Becker R.~H., Leighly K.~M., 2001, AJ, 121, 2915 
\bibitem[\protect\citeauthoryear{Blanton et al.}{2003}]{Blanton03} 
        Blanton E.~L., Gregg M.~D., Helfand D.~J., Becker R.~H., White R.~L., 2003, AJ, 125, 1635 
\bibitem[\protect\citeauthoryear{B{\"o}hringer et al.}{2001}]{Boehringer01} 
        B{\"o}hringer H., et al., 2001, A\&A, 369, 826 
\bibitem[\protect\citeauthoryear{Borgani et al.}{2004}]{Borgani04} 
        Borgani S., et al., 2004, MNRAS, 348, 1078 
\bibitem[\protect\citeauthoryear{Brown \& Rudnick}{2009}]{Brown09} 
        Brown S., Rudnick L., 2009, AJ, 137, 3158
\bibitem[\protect\citeauthoryear{Brunetti et al.}{2007}]{Brunetti07}
        Brunetti G., Venturi T., Dallacasa D., Cassano R., Dolag K., Giacintucci S., Setti G., 2007, ApJ, 670, L5
\bibitem[\protect\citeauthoryear{Burns}{1990}]{Burns90} 
        Burns J.~O., 1990, AJ, 99, 14 
\bibitem[\protect\citeauthoryear{Burns}{1998}]{Burns98} 
        Burns J.~O., 1998, Science, 280, 400
\bibitem[\protect\citeauthoryear{Cassano}{2009}]{Cassano09} 
        Cassano R., 2009, in The Low-Frequency Radio Universe, eds Saikia D.J., Green D.A., Gupta Y, Venturi T.,
        ASPC, 407, 223
\bibitem[\protect\citeauthoryear{Chiaberge et al.}{2009}]{Chiaberge09} 
        Chiaberge M., Tremblay G., Capetti A., Macchetto F.~D., Tozzi P., Sparks W.~B., 2009, ApJ, 696, 1103 
\bibitem[\protect\citeauthoryear{Colless et al.}{2001}]{Colless01} 
        Colless M., et al., 2001, MNRAS, 328, 1039 
\bibitem[\protect\citeauthoryear{Croston et al.}{2004}]{Croston04} 
        Croston J.~H., Birkinshaw M., Hardcastle M.~J., Worrall D.~M., 2004, MNRAS, 353, 879 
\bibitem[\protect\citeauthoryear{Croston et al.}{2005}]{Croston05} 
        Croston J.~H., Hardcastle M.~J., Harris D.~E., Belsole E., Birkinshaw M., Worrall D.~M., 2005, ApJ, 626, 733 
\bibitem[\protect\citeauthoryear{De Lucia \& Blaizot}{2007}]{DeLucia07} 
        De Lucia G., Blaizot J., 2007, MNRAS, 375, 2
\bibitem[\protect\citeauthoryear{En{\ss}lin et al.}{1998}]{Ensslin98} 
        En{\ss}lin T.~A., Biermann P.~L., Klein U., Kohle S., 1998, A\&A, 332, 395
\bibitem[\protect\citeauthoryear{En{\ss}lin \& Gopal-Krishna}{2001}]{Ensslin01}
        En{\ss}lin T.~A., Gopal-Krishna, 2001, 366, 26 
\bibitem[\protect\citeauthoryear{Fanaroff \& Riley}{1974}]{Fanaroff74} 
        Fanaroff B.~L., Riley J.~M., 1974, MNRAS, 167, 31P 
\bibitem[Feretti(2005)]{Feretti05} 
        Feretti L., 2005, AdSpR, 36, 729
\bibitem[\protect\citeauthoryear{Ferrari et al.}{2008}]{Ferrari08} 
        Ferrari C., Govoni F., Schindler S., Bykov A.~M., Rephaeli Y., 2008, SSRv, 134, 93 
\bibitem[\protect\citeauthoryear{Giacintucci \& Venturi}{2009}]{Giacintucci09}
        Giacintucci S., Venturi T., 2009, A\&A, 505, 55 
\bibitem[\protect\citeauthoryear{Giacintucci et al.}{2007}]{Giacintucci07} 
        Giacintucci S., Venturi T., Murgia M., Dallacasa D., Athreya R., Bardelli S., Mazzotta P., Saikia D.J., 2007, A\&A, 476, 99 
\bibitem[\protect\citeauthoryear{Giacintucci et al.}{2008}]{Giacintucci08} 
        Giacintucci S., et al., 2008, A\&A, 486, 347 
\bibitem[Giovannini \& Feretti(2000)]{Giovannini00} 
        Giovannini G., Feretti L., 2000, NewA, 5, 335
\bibitem[Giovannini \& Feretti(2004)]{Giovannini04} 
        Giovannini G., Feretti L., 2004, JKAS, 37, 323 
\bibitem[\protect\citeauthoryear{Giovannini et al.}{2009}]{Giovannini09} 
        Giovannini G., Bonafede A., Feretti L., Govoni F., Murgia M., Ferrari F., Monti G., 2009, A\&A, 507, 1257
\bibitem[\protect\citeauthoryear{Harris, Kapahi \& Ekers}{1980}]{Harris80}
        Harris D.E., Kapahi V.K., Ekers R.D., 1980, A\&AS, 39, 215 
\bibitem[\protect\citeauthoryear{Harris et al.}{1993}]{Harris93}
        Harris D.~E., Stern C.~P., Willis A.~G., Dewdney P.~E., 1993, AJ, 105, 769
\bibitem[\protect\citeauthoryear{Hoeft et al.}{2008}]{Hoeft08} 
        Hoeft M., Br{\"u}ggen M., Yepes G., Gottl{\"o}ber S., Schwope A., 2008, MNRAS, 391, 1511 
\bibitem[\protect\citeauthoryear{Johnston-Hollitt}{2003}]{mjh03} 
        Johnston-Hollitt M., 2003, PhD Thesis, University of Adelaide
\bibitem[\protect\citeauthoryear{Johnston-Hollitt, Hunstead \& Corbett}{2008}]{mjh08} 
        Johnston-Hollitt M., Hunstead R.~W., Corbett E., 2008, A\&A, 479, 1 
\bibitem[\protect\citeauthoryear{Johnston-Hollitt et al.}{2010}]{mjh10} 
        Johnston-Hollitt M., Gill J.~A., Hollitt C.~P., Tunstall L. 2010, MNRAS, submitted
\bibitem[\protect\citeauthoryear{Kantharia, Das \& Gopal-Krishna}{2009}]{Kantharia09}
        Kantharia N.G., Das M., Gopal-Krishna, 2009, JApA, 30, 37 
\bibitem[\protect\citeauthoryear{Konar et al.}{2004}]{Konar04} 
        Konar C., Saikia D.~J., Ishwara-Chandra C.~H., Kulkarni V.~K., 2004, MNRAS, 355, 845 
\bibitem[\protect\citeauthoryear{Konar et al.}{2009}]{Konar09} 
        Konar C., Hardcastle M.~J., Croston J.~H., Saikia D.~J., 2009, MNRAS, 400, 480
\bibitem[\protect\citeauthoryear{Kravtsov et
          al.}{2009}]{Kravtsov09} Kravtsov A., et al., 2009, astro,
        2010, 164
\bibitem[\protect\citeauthoryear{Lanzetta et al.}{2002}]{Lanzetta02} 
        Lanzetta K.~M., Yahata N., Pascarelle S., Chen H.-W., Fern{\'a}ndez-Soto A., 2002, ApJ, 570, 492 
\bibitem[\protect\citeauthoryear{Lewis et al.}{2002}]{Lewis02} 
        Lewis I.~J., et al., 2002, MNRAS, 333, 279 
\bibitem[\protect\citeauthoryear{Lonsdale et al.}{2003}]{Lonsdale03} 
        Lonsdale C.~J., et\,al., 2003, PASP, 115, 897 
\bibitem[\protect\citeauthoryear{Mao et al.}{2009a}]{Mao09a} 
        Mao M.~Y., Johnston-Hollitt M., Stevens J.~B., Wotherspoon S.~J., 2009a, MNRAS, 392, 1070 
\bibitem[\protect\citeauthoryear{Mao et al.}{2009b}]{Mao09b} 
        Mao M.~Y., Norris R.~P., Sharp R., Lovell J.~E.~J., 2009b, ASPC, 408, 380 
\bibitem[\protect\citeauthoryear{Matthews, Morgan \& Schmidt}{1964}]{Matthews64} 
        Matthews T.~A., Morgan W.~W., Schmidt M., 1964, ApJ, 140, 35 
\bibitem[\protect\citeauthoryear{Middelberg et al.}{2008}]{Middelberg08} 
        Middelberg E., et al., 2008, AJ, 135, 1276 
\bibitem[\protect\citeauthoryear{Miniati et al.}{2000}]{Miniati00} 
        Miniati F., Ryu D., Kang H., Jones T.~W., Cen R., Ostriker J.~P., 2000, ApJ, 542, 608 
\bibitem[\protect\citeauthoryear{Miszalski et al.}{2006}]{Miszalski06} 
        Miszalski B., Shortridge K., Saunders W., Parker Q.~A., Croom S.~M., 2006, MNRAS, 371, 1537 
\bibitem[\protect\citeauthoryear{Norris et al.}{2006}]{Norris06} 
        Norris R.~P., et\,al., 2006, AJ, 132, 2409 
\bibitem[\protect\citeauthoryear{Norris et al.}{2009}]{Norris09} 
        Norris R.~P., the EMU team, 2009, PoS(PRA2009) 033
\bibitem[\protect\citeauthoryear{Oklopcic et
          al.}{2010}]{Oklopcic10} Oklop{\v c}i{\'c} A., et al., 2010,
        ApJ, 713, 484
\bibitem[\protect\citeauthoryear{Owen \& Rudnick}{1976}]{Owen76} 
        Owen F.~N., Rudnick L., 1976, ApJ, 205, L1 
\bibitem[\protect\citeauthoryear{Owen \& Ledlow}{1994}]{Owen94} 
        Owen F.~N., Ledlow M.~J., 1994, ASPC, 54, 319 
\bibitem[\protect\citeauthoryear{Owen \& Ledlow}{1997}]{Owen97} 
        Owen F.~N., Ledlow M.~J., 1997, ApJS, 108, 41
\bibitem[\protect\citeauthoryear{Owers, Couch \& Nulsen}{2009}]{Owers09} 
        Owers M.~S., Couch W.~J., Nulsen P.~E.~J., 2009, ApJ, 693, 901 
\bibitem[\protect\citeauthoryear{Pfrommer et al.}{2006}]{Pfrommer06} 
        Pfrommer C., Springel V., En{\ss}lin T.~A., Jubelgas M., 2006, MNRAS, 367, 113 
\bibitem[\protect\citeauthoryear{Pfrommer, En{\ss}lin \& Springel}{2008}]{Pfrommer08} 
        Pfrommer C., En{\ss}lin T.~A., Springel V., 2008, MNRAS, 385, 1211 
\bibitem[\protect\citeauthoryear{Pinkney et al.}{2000}]{Pinkney00} 
        Pinkney J., Burns J.~O., Ledlow M.~J., G{\'o}mez P.~L., Hill J.~M., 2000, AJ, 120, 2269 
\bibitem[\protect\citeauthoryear{Pratt et al.}{2009}]{Pratt09} 
        Pratt G.~W., Croston J.~H., Arnaud M., B{\"o}hringer H., 2009, A\&A, 498, 361 
\bibitem[\protect\citeauthoryear{Ricker \& Sarazin}{2001}]{Ricker01} 
        Ricker P.~M., Sarazin C.~L., 2001, ApJ, 561, 621
\bibitem[\protect\citeauthoryear{Roettiger, Burns \& Stone}{1999}]{Roettiger99}
        Roettiger K., Burns J.O., Stone J.M., 1999, ApJ, 518, 603 
\bibitem[Rowan-Robinson et al.(2008)]{Rowan-Robinson08} 
        Rowan-Robinson, M., et al.\ 2008, MNRAS, 386, 697 
\bibitem[\protect\citeauthoryear{Rudnick \& Owen}{1977}]{Rudnick77}
        Rudnick L., Owen F.N., 1977, AJ, 82, 1 
\bibitem[\protect\citeauthoryear{Ryu et al.}{2003}]{Ryu03} 
        Ryu D., Kang H., Hallman E., Jones T.~W., 2003, ApJ, 593, 599 
\bibitem[\protect\citeauthoryear{Sahl\'en et al.}{2009}]{Sahlen09} 
        Sahl\'en M., et al., 2009, MNRAS, 397, 577  
\bibitem[\protect\citeauthoryear{Saikia \& Jamrozy}{2009}]{Saikia09} 
        Saikia D.~J., Jamrozy M., 2009, BASI, 37, 63 (arXiv:1002.1841)
\bibitem[\protect\citeauthoryear{Saikia, Wiita \& Muxlow}{1993}]{Saikia93} 
        Saikia D.~J., Wiita P.~J., Muxlow T.~W.~B., 1993, AJ, 105, 1658 
\bibitem[\protect\citeauthoryear{Saikia et al.}{1987}]{Saikia87} 
Saikia D.~J., Staveley-Smith L., Wills D., Cornwell T.~J., Salter C.~J., 
Junor W., Shastri P., 1987, MNRAS, 229, 495 
\bibitem[\protect\citeauthoryear{Sarazin}{1999}]{Sarazin99} 
        Sarazin C.~L., 1999, ApJ, 520, 529 
\bibitem[\protect\citeauthoryear{Saunders et al.}{2004}]{Saunders04} 
        Saunders W., et\,al., 2004, SPIE, 5492, 389 
\bibitem[\protect\citeauthoryear{Schwartz et al.}{2006}]{Schwartz06} 
        Schwartz D.~A., et al., 2006, ApJ, 640, 592 
\bibitem[\protect\citeauthoryear{Sharp et al.}{2006}]{Sharp06} 
        Sharp R., et\,al., 2006, SPIE, 6269, 14 
\bibitem[\protect\citeauthoryear{Smolci\'c et al.}{2006}]{Smolcic07}
        Smolci\'c V., et al., 2007, ApJS, 172, 295 
\bibitem[\protect\citeauthoryear{Tribble}{1993}]{Tribble93}
        Tribble P.~C., 1993, MNRAS, 263, 31 
\bibitem[\protect\citeauthoryear{Vazza, Brunetti \& Gheller}{2009}]{Vazza09} 
        Vazza F., Brunetti G., Gheller C., 2009, MNRAS, 395, 1333 
\bibitem[\protect\citeauthoryear{Venturi et al.}{2008}]{Venturi08}
        Venturi T., Giacintucci S., Dallacasa D., Cassano R., Brunetti G., Bardelli S., Setti G., 2008, A\&A, 484, 327 
\bibitem[\protect\citeauthoryear{Wright}{2006}]{Wright06} 
        Wright E.~L., 2006, PASP, 118, 1711

\end{thebibliography}
\end{document}